\documentclass[preprint,12pt]{elsarticle}
\pdfoutput=1
\usepackage{amssymb}
\usepackage{amsthm}
\usepackage{amsmath}
\usepackage{hyperref}
\usepackage{caption}
\usepackage{subcaption}
\usepackage{colortbl}
\usepackage{booktabs}
\usepackage{verbatim}
\usepackage{lineno}
\usepackage{color}
\usepackage{multirow}
\usepackage{graphicx} 
\graphicspath{{figures/}}
\newcommand{\etal}{et al.}

\journal{Arxiv}
\begin{document}
\begin{frontmatter}
\title{Stochastic Porous Microstructures}
\affiliation[inst1]{
   organization={School of Computer Science and Technology, Shandong University},
    city={Qingdao},
    country={China}
}
\affiliation[inst2]{
    organization={Computer Science Department, Ben-Gurion University},
    city={Beer-Sheva},
    country={Israel}
}
\author[inst1]{Zhongren Wang}
\author[inst1]{Lihao Tian}
\author[inst1]{Xiaokang Liu}
\author[inst2]{Andrei Sharf}
\author[inst1]{Lin Lu\texorpdfstring{\corref{mycorrespondingauthor}}{}}
\cortext[mycorrespondingauthor]{Corresponding author}
\ead{llu@sdu.edu.cn}
\begin{abstract}
Stochastic porous structures are ubiquitous in natural phenomena and have gained considerable traction across diverse domains owing to their exceptional physical properties. The recent surge in interest in microstructures can be attributed to their impressive attributes, such as a high strength-to-weight ratio, isotropic elasticity, and bio-inspired design principles. Notwithstanding, extant stochastic structures are predominantly generated via procedural modeling techniques, which present notable difficulties in representing geometric microstructures with periodic boundaries, thereby leading to intricate simulations and computational overhead.
In this manuscript, we introduce an innovative method for designing stochastic microstructures that guarantees the periodicity of each microstructure unit to facilitate homogenization. We conceptualize each pore and the interconnecting tunnel between proximate pores as Gaussian kernels and leverage a modified version of the minimum spanning tree technique to assure pore connectivity.
We harness the dart-throwing strategy to stochastically produce pore locations, tailoring the distribution law to enforce boundary periodicity. We subsequently employ the level-set technique to extract the stochastic microstructures. Conclusively, we adopt Wang tile rules to amplify the stochasticity at the boundary of the microstructure unit, concurrently preserving periodicity constraints among units.
Our methodology offers facile parametric control of the designed stochastic microstructures. Experimental outcomes on 3D models manifest the superior isotropy and energy absorption performance of the stochastic porous microstructures. We further corroborate the efficacy of our modeling strategy through simulations of mechanical properties and empirical experiments.
\end{abstract}
\begin{keyword}
porous structure \sep stochastic microstructure \sep boundary periodicity \sep bicontinuous structure
\end{keyword}
\end{frontmatter}

\section{Introduction}
Porous structures are widely found in nature, such as natural clinoptilolites, biological bones, beehives, butterfly fins, etc.
Such structures possess excellent physical properties, such as high stiffness-to-weight ratio, permeability, and controllable Poisson's ratio~\cite{Gibson1997}, and thus are widely used in various applications, such as bone substitutes in medicine, catalyst carrier in chemical engineering, or shock-absorbing structures in the industry. 
In line with the rapid development of additive manufacturing techniques that enable the production of porous structures with high geometric complexity, modeling methods for exploiting the capabilities of high geometric freedom porous structures with desired physical properties, such as a high strength-to-weight ratio, has become a popular research topic.
To meet the manufacturing and functional requirements, the structural design needs to consider 3D printing constraints and high-performance requirements for specific applications.
For example, the porous structure needs to have connected internal cavities instead of isolated closed pores to ensure the material discharge process of 3D printing techniques; the smooth surface may reduce stress concentration and thus enhance mechanical properties.

Stochastic porous structures, emulating the randomness and irregularity found in natural porous materials, distinguish themselves from various kinds of porous structures through their unique attributes. For instance, the inherent randomness of these structures facilitates the achievement of isotropy in their physical properties, a vital characteristic often sought after in microstructure applications.
Therefore modeling stochastic porous structures has attracted much interest in the computational fabrication community.
Stochastic porous structures have multiple modeling methods. A large class of methods is based on Voronoi cells, and different Voronoi tessellations are obtained through the distribution of points to achieve the stochasticity of the structure~\cite{Lu2014, Martinez2016, Kou2010}.
In addition, stochastic porous structures with different geometric features can also be achieved through different implicit functions.
Al-Ketan \etal~\cite{alketanmechanical2021} design partially closed sheet-based stochastic porous structures based on Gyroid structure and emphasize the isotropy of stochastic porous structures.
Tian \etal~\cite{Tian2020} designed an organic porous structure with stochasticity while considering the manufacturing constraints to ensure the structure is biomimetic.
In specific applications, the stochastic porous structure design approach involves using the entire internal region of the model as the design space; therefore it requires large data occupancy and a computationally costly mechanical simulation process.
Besides, design approaches of stochastic porous structures usually require the identification of some physical performance requirement in the specific application and then obtaining a fine-grained final structure through optimization methods which also requires a large amount of calculation, therefore becomes the bottleneck of stochastic structure design and application.

In order to overcome the difficulty of the computational cost in the modeling process of porous structures, the natural idea is to design periodic microstructures to reduce the amount of computation and representation.
Periodic microstructures, also known as metamaterials, have units with repeating boundaries.
In specific applications, the periodic microstructures fill the entire model by tessellation, and the unit length is much smaller than the scale of the model.
With this scale variation, the periodic microstructure can be regarded as a special new material with uniform physical properties.
The homogenization method can calculate the equivalent physical properties of periodic microstructures.
A material property space is constructed by changing the geometric parameters required in the design.
In this periodic microstructure design approach, the structural design space is limited to the size of microstructure units, which can significantly reduce the amount of data required to express the whole complex porous structures. 
For example, common porous microstructure units include body-center cubic (BCC) \cite{Maskery2016} and face-centered cubic (FCC) \cite{Ahmadi2014}, as well as triply periodic minimal surfaces (TPMS) \cite{Yan2020,Hu2022}.
But as periodic unit cells, usually only a single microstructure is used within the overall structure, which is difficult to achieve continuous variation of microstructure design parameters and properties in one model; therefore, the application scenarios of porous structures are constrained.

To summarize, stochastic microstructures and periodic microstructures each possess different desirable properties. Our insight is to develop a novel modeling approach that combines the advantages of both types of microstructures to design and fabricate porous structures with connected, open pores and smooth surfaces, while adhering to the constraints of 3D printing. Our objective is to strike a balance between the global stochasticity of porosity and the computational performance of the periodic microstructure unit. In achieving this goal, we face critical challenges such as resolving the natural conflict between stochasticity and periodicity, reducing the size of the design domain to save data expression, choosing appropriate geometric expressions, and constructing a parametric design method.

In this paper, we present a systematic framework for designing stochastic and periodic porous microstructures. First, under the concept of periodic microstructures, we design unit cells of porous structures and periodic boundaries to ensure that the unit cells are splicable. We then tile cells with the same boundaries to maintain structural continuity and achieve periodicity, while introducing stochasticity within the cells. Specifically, we follow the organic pore structure representation~\cite{Tian2020}, which considers each pore as a Gaussian kernel. The periodicity of the boundary is ensured by controlling the pore distribution on the boundary through our algorithm. Finally, inspired by the traditional Wang tile~\cite{Wang1961}, we select a series of unit cells with different inner porous structures and fixed boundaries, and then combine the cells according to the Wang tile rules to manufacture porous microstructures with higher stochasticity.

The paper introduces a modeling framework for designing stochastic microstructures with boundary periodicity and bi-continuity. The stochasticity is achieved not only from the interior of the microstructure unit but also from among the neighboring units through customized Wang tile rules. The proposed framework offers physical properties verification of stochastic microstructure in different cell tiling levels and the functionally gradient structure. Overall, the contributions of this paper include the development of a flexible and effective approach for generating complex stochastic microstructures and providing a physical basis for their properties.

Our contributions are summarized as follows:
\begin{itemize}
    \item We propose a novel, generalized methodology for the modeling of interconnected, stochastic porous structures, thereby pushing the boundaries of conventional porous structure design.
    \item We introduce a scalable approach leveraging the principles of Wang Tile, preserving key properties such as stochasticity, porosity, and connectivity within the generated structures.
    \item We guarantee the periodicity of microstructures to facilitate homogenization.
    \item We devise a methodology for direct control over porosity and connectivity, providing a tangible pathway for the precise design and optimization of porous structures according to specific application requirements.
\end{itemize}
\section{Methodology}
\label{sec:method}
In Section~\ref{subsec:periodic}, we will describe the method for modeling stochastic periodic porous microstructures. In Section~\ref{subsec:stochastic}, we will introduce the global stochasticity method, while Section~\ref{subsec:porosity} will be dedicated to discussing porosity control. Finally, in Section~\ref{subsec:meshfree}, we will present an efficient method for generating manufacturing files based on mesh-free modeling.

\begin{figure}[htp]
\centering
        \includegraphics[width=.24\textwidth]{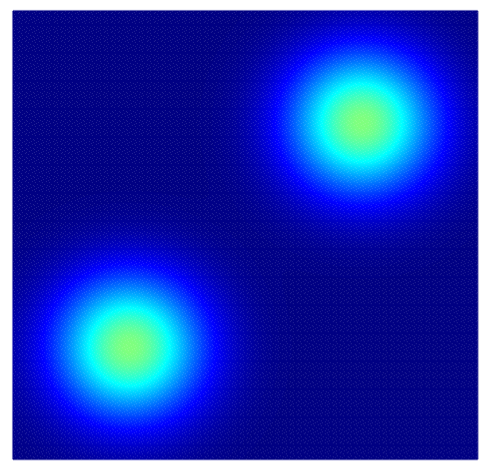}
        \includegraphics[width=.24\textwidth]{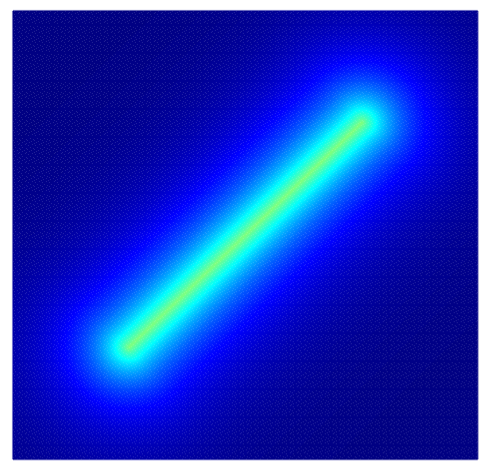}
        \includegraphics[width=.24\textwidth]{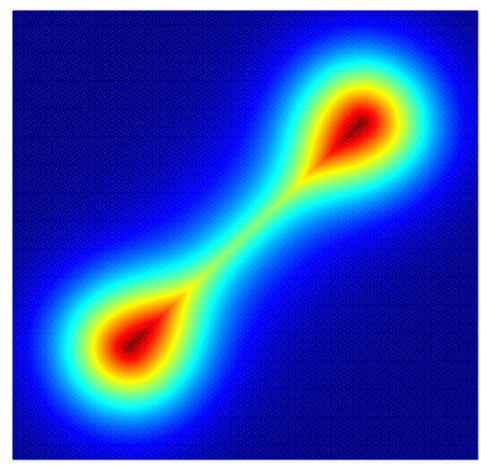}
        \includegraphics[width=.24\textwidth]{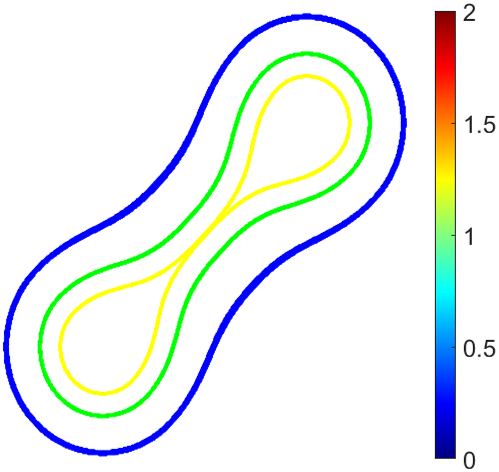}
        \centerline{(a)\hspace{0.2\linewidth}(b)\hspace{0.2\linewidth}(c)\hspace{0.2\linewidth}(d)}
\caption{Visualization results of the pore function~(a), tunnel function~(b), combined function~(c) in two dimensions and three extracted level-set curves from combined function~(d). The corresponding level-set values for blue, green and yellow are 0.2, 0.6 and 1.0, respectively.
    }
    \label{fig:Combined function2d}
\end{figure}

\subsection{Stochastic Periodic Porous Microstructure (SPPM) Modeling}
\label{subsec:periodic}

\textbf{Pore and tunnel modelling.} 
Because experimental results show that the porous structure generated by the Gaussian kernel function has a smooth boundary and can offer biomimetic properties, we regard the pore as a Gaussian kernel.
~Specifically, $p_i(x_i,y_i,z_i)$ is the position of the center of the pore, and the basic 3D isotropic Gaussian kernel function is defined as
\begin{equation}
  Gp_i(p) = e^{-\omega_i{\left \| p-p_i \right \|^2}}, ~~p\in \Omega 
\end{equation}
where $\omega_i > 0 $  determines the size of the pore, $p$ is the sample point in the three-dimensional space $\Omega$. 
Fig.~\ref{fig:Combined function2d}(a) illustrates the function of two pores in two dimensions.

To make the generated porous structure with internal connectivity, we propose an implicit method for representing tunnels, which smoothly combines with the representation of pores.
Specifically, we define the structure connecting two pores $p_i$ and $p_j$ as a "tunnel" $t_{ij}$.
We consider the tunnel $t_{ij}$ connecting the pores $p_i$ and $p_j$ as an edge $s_{ij}$ between the center of $p_i$ and $p_j$.
The tunnel function can be expressed as a transformed Gaussian function by taking the edge as the Gaussian kernel:
\begin{equation}
  Gt_{ij}(p) = e^{-\mu _{ij}{\left \| p-s_{ij} \right \|}}, ~~p\in \Omega, 
\end{equation}
where $\mu  _{ij} > 0$ controls the minimum radius of the tunnel.
$\left \| \cdot  \right \|$ can be seen as the distance from a point to a line segment. 
The result of the tunnel function visualized in two dimensions is shown in Fig.~\ref{fig:Combined function2d}(b).

To combine the pore and tunnel functions, we create a combined function to represent the porous structure. 
Assuming that there is a set of pores $P = \left \{ p_1\dots p_n\right \} \subset \mathbb{R}^3 $ with the number of pores $n$, we define the combined function $\mathcal{G}$ for every point $p \in \Omega$ and every tunnel $t_{ij}$ as 
\begin{equation}
\mathcal{G}(p)  = \sum_{i = 1}^{n}  Gp_i(p) + \sum_{}^{} Gt_{ij}(p).
\end{equation}

The result of the combined function visualized in two dimensions is shown in Fig.~\ref{fig:Combined function2d}(c). 

The parametric model of SPPM is defined as,

\begin{equation}
\label{equa:SPPM Modeling}
\begin{aligned} 
    M &= G(P, \theta), \\
    P &= \left ( p_i \right )_{i=1}^n, \\
    \theta  &= \left \{ \omega ,\mu  \right \}.
\end{aligned}
\end{equation}

The stochastic periodic porous microstructure (SPPM), denoted as $M$, is defined using two sets of parameters: the pore distribution parameters, $P$, and the pore generation parameters, $\theta$. $P = \left ( p_i \right )_{i=1}^n$ signifies the spatial arrangement of the pores, where $n$ indicates the total count of pores and $p_i$ marks the location of each individual pore. The pore generation parameters, $\theta$, encompass the pore size parameter $\omega$, and the tunnel size parameter $\mu$. These parameters profoundly influence both the geometrical configuration and the physical attributes of the SPPM. Initially, we employ the Gaussian kernel function to implicitly represent the pore and tunnel, where $\omega$ specifies the radius of the pore and $\mu$ determines the tunnel's minimum radius. Subsequently, we employ a blue noise sampling method to optimize the distribution of the pores. We impose a boundary periodicity rule to regulate the distribution of the outermost pores. Once $\omega$ and $\mu$ are established, we fine-tune the number of pores, $n$, to attain the desired porosity. The signed distance field (SDF) is built on the basis of the parametric model $M$. We then resort to the level-set method to extract the porous microstructure's surface, enabling us to create a mesh or voxel model. The forthcoming sections will delve into the details of the SPPM modeling process.

\begin{figure}[htp]
    \centering
    \includegraphics[width=.6\linewidth]{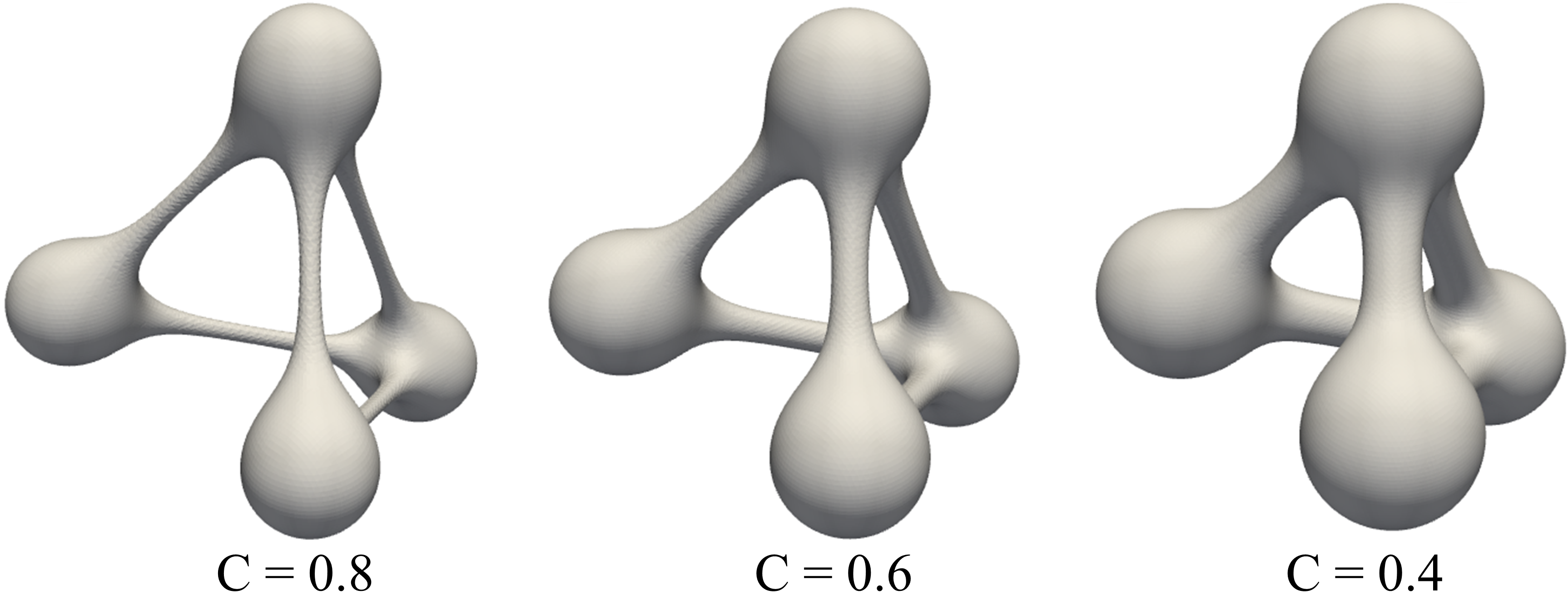}
    \caption{The porous structures under different level-set values.}
    \label{fig:Combined function3d}
\end{figure}

Now that we have the implicit representation of pores and tunnels, the next step is to extract the structure surface by the implicit representation using the level-set method.
The level-set method is a numerical technique for using level sets as a tool for the numerical analysis of surfaces and shapes. 
The smooth boundary of solid and void parts can be represented by a constant value $C$ in an implicit function, which in our work can be represented as
\begin{equation}
\begin{cases}\mathcal{G}(p) = C ,p \in \partial \Omega 
 \\\mathcal{G}(p) < C,p\in \Omega ^-
 \\\mathcal{G}(p) > C,p\in \Omega ^+
\end{cases}
\end{equation}
where $\mathcal{G}$ is a level-set function at sample point $p$ which are the coordinates of the points in three-dimensional space.
$\partial \Omega $ is the extracted surface, $\Omega ^-$ is the interior of the structure, and $\Omega ^+$ is the exterior of the structure.
The curve of the combined function for different level set values is shown in Fig.~\ref{fig:Combined function2d}(d).
Fig.~\ref{fig:Combined function3d} shows the surface of the porous structure obtained in 3D space by taking different level-set values. Due to the nature of implicit functions, our parametric approach guarantees a smooth structure.

As mentioned above, we designed a uniform set of implicit representations of pores and tunnels based on the Gaussian kernel function.

\begin{figure}[htp]
\centering
    \includegraphics[width=0.3\textwidth]{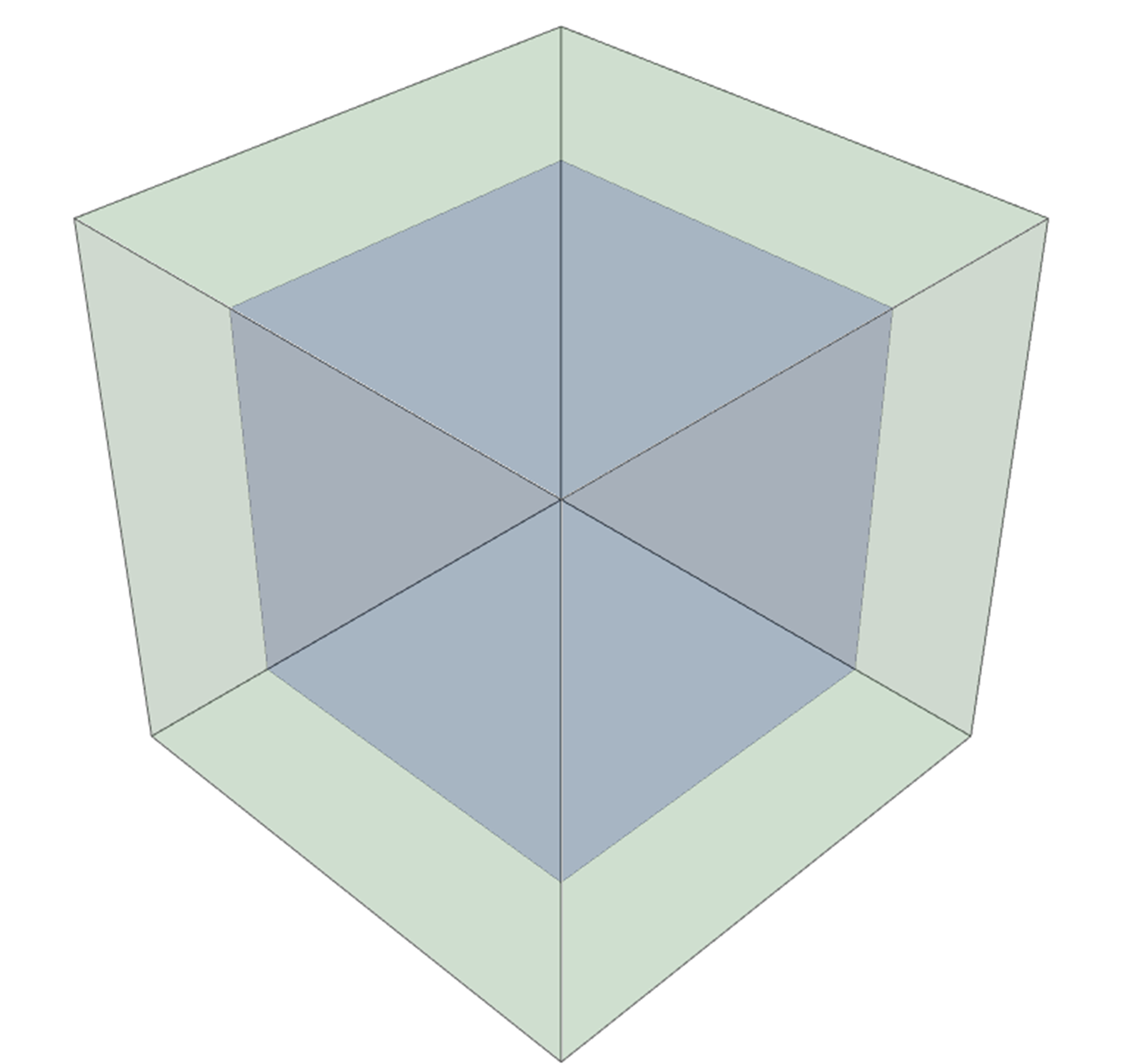}
    \includegraphics[width=0.3\textwidth]{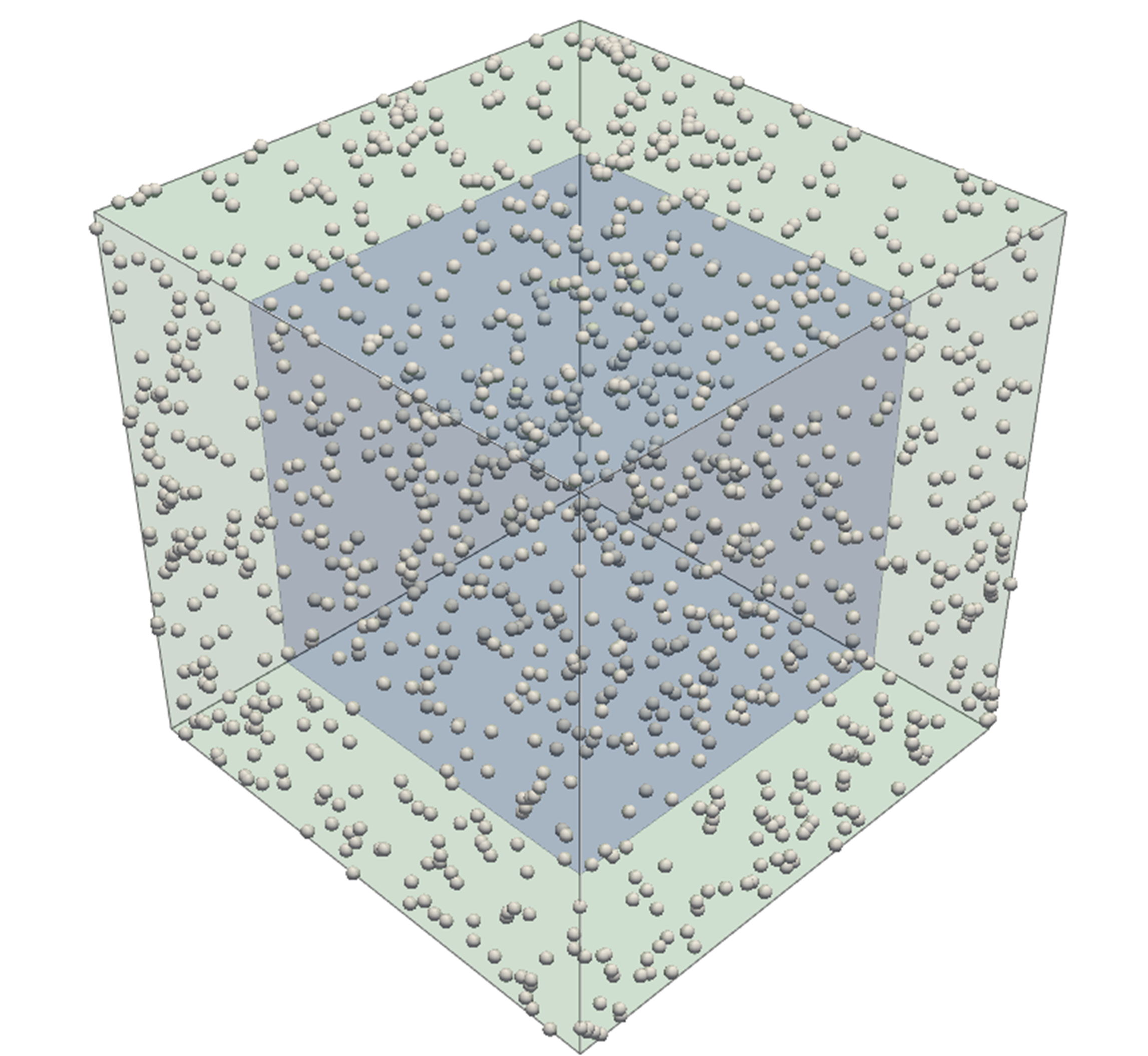}
    \includegraphics[width=.3\textwidth]{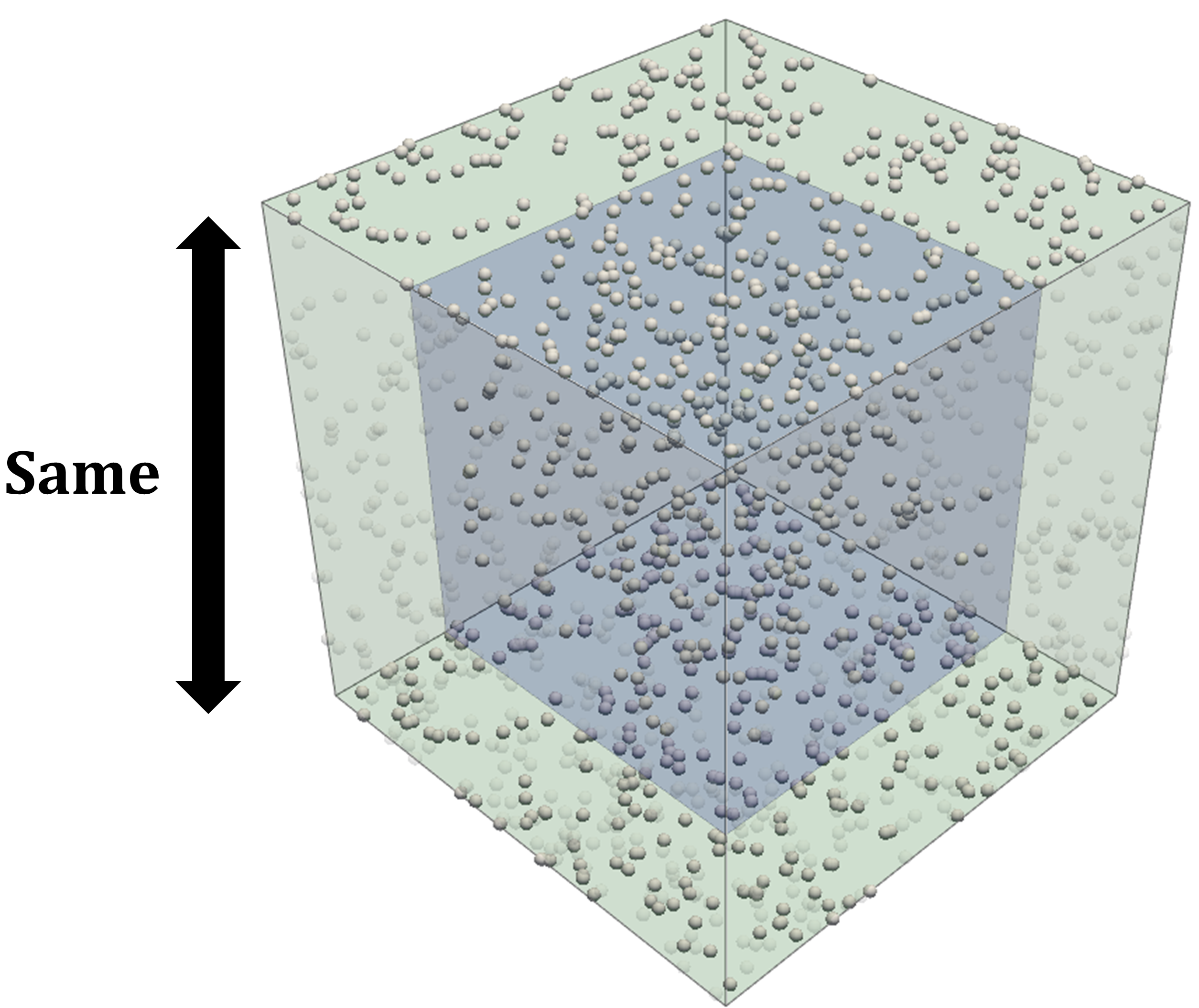}
\centerline{(a)\hspace{0.28\linewidth}(b)\hspace{0.28\linewidth}(c)}    
    \caption{Region division of microstructure. (a) Dividing the porous microstructure into two regions: interior indicated in blue and surface indicated in green. (b) Individual pores generation in the interior and on the surface. (c) Consistent distribution of pores on two opposing surfaces.}
    \label{fig:pore distribution}
\end{figure}

\textbf{Boundary Periodicity Regulation.}
The objective of implementing boundary periodicity is to preserve the coherence among contiguous microstructures throughout the tessellation procedure. Nonetheless, the typical implication of boundary periodicity is the uniformity of topology or distribution, subsequently compromising the macroscopic stochasticity of the microstructure.

In order to strike a balance between boundary periodicity and the comprehensive stochasticity of the microstructure, we have devised a rule to supervise the distribution of pores. A key precondition for boundary periodicity is the mirror symmetry of the solid components on juxtaposed surfaces. Given that our microstructure is ortho-hexahedral, we generate mirror-symmetric pores on two diametrically opposed surfaces for an SPPM. To facilitate the aforementioned operation, we segregate the modeling space of SPPM into two domains: the surface and the interior. The distance from the surface to the interior, denoted by $d$, is determined based on the predefined porosity. As depicted in Fig.~\ref{fig:pore distribution}(a), the interior region is delineated as a regular hexahedron with a side length less than the unit, represented in blue, with the unit surface marked in green. Pores are simultaneously generated on both the surface and interior. Fig.~\ref{fig:pore distribution}(c) elucidates the pore distribution regulations of a porous microstructure, in other words, diametric surfaces maintain identical pore distribution.

\textbf{Stochasticity of SPPM.}
To harness the advantageous trait of stochasticity, we employ a stochastic method to ascertain the locations of the pores. It is crucial to avoid the occurrence of fusion phenomena caused by the overly close proximity between any two pores, which can result in a localized decrement in material density and weakened connectivity. To circumvent the issue of excessively low local rigidity, which could render the structure incapable of satisfying functional prerequisites due to this stochasticity, we must regulate the distance between every two pores.

Consequently, we utilize a blue noise sampling technique to manage the spacing of the pores. More specifically, upon the generation of the subsequent pore, if the distance from that pore to all previously generated pores exceeds a predefined distance threshold $l$, the pore is retained; on the contrary, the pore is discarded and reformed until the quantity of generation achieves the predetermined number. This pore-generation strategy is executed separately on the surface and interior.

\begin{figure}[htp]
        \centering
        \includegraphics[width=0.3\textwidth]{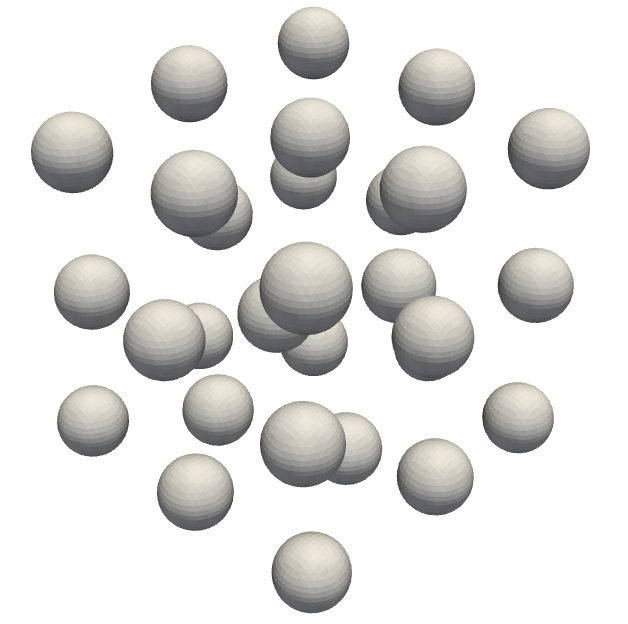}
        \includegraphics[width=0.3\textwidth]{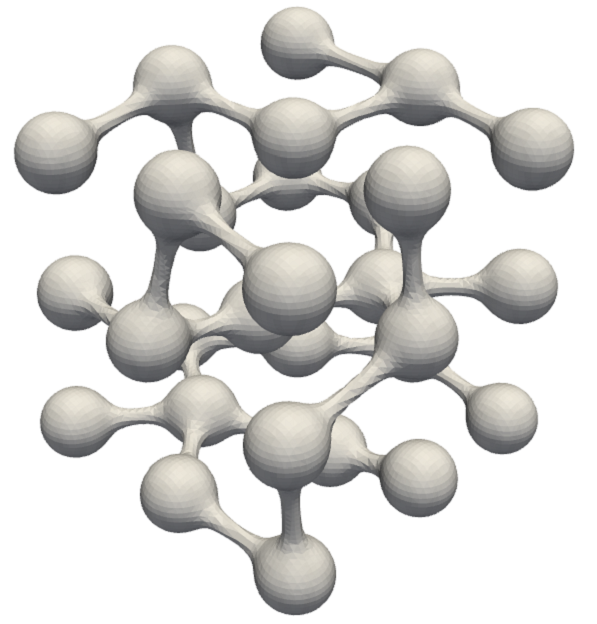}
        \includegraphics[width=0.3\textwidth]{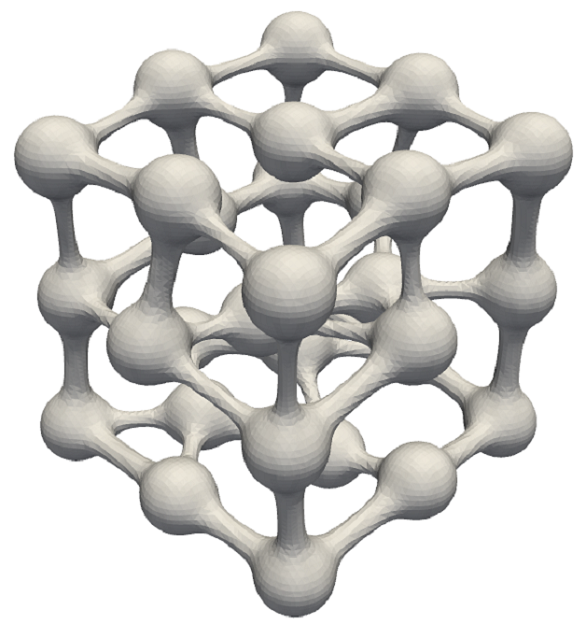}
\centerline{(a)\hspace{0.28\linewidth}(b)\hspace{0.28\linewidth}(c)}        
    \caption{Comparison between MBDST and traditional MST methods on 30 pores. (a) 30 pores without connections; (b) the MST connection result; (c) the MBDST connection result with the degree lower bound 3 and degree upper bound 5.}
    \label{fig:MST}
\end{figure}

\textbf{Reinforcement of Pore Connectivity.} In the pursuit of generating tunnels with an adjustable level of connectivity between pores, we employ the minimum bound degree spanning tree method (MBDST)~\cite{Singh2015}.
We define this challenge by contemplating all generated pores as the vertex set $V$ in an undirected graph $G$ and the tunnels as the edge set $E$ within this undirected graph.
To augment the connectivity amidst the pores while concurrently averting the interlocking between the formulated tunnels, we establish a degree lower bound $A_v$ and an upper bound $B_v$. The objective is to identify a minimal cost that gratifies both the upper and lower degree constraints. We achieved pore connectivity on 30 pores, employing the conventional MST method and the MBDST with $A_v = 3$, $B_v = 5$, correspondingly.
Fig.~\ref{fig:MST} exhibits the comparison between the application of the MBDST method and the conventional MST method on 30 pores.
The observation is that the structure derived through the MBDST method exhibits superior connectivity.
The MBDST method is separately and concurrently applied on the surface and the interior, ultimately generating a tunnel between each pore on the surface and its closest counterpart on the interior.

\textbf{Solid Structure Connectivity Verification.} While the interconnection among pores is assured, we need to validate the connectivity within the solid structures. Due to the stochasticity of the pore location, the solid structure is likely not a connected component. To address this issue, our approach retains the largest connected component of the solid structure as an SPPM. Nevertheless, if there are solitary solid structures at the surface, this strategy can potentially compromise the boundary periodicity of the microstructure. To uphold the boundary periodicity of the SPPM, we designate the SPPM generated to be an erroneous result if the deleted connected component is situated at the surface. Our generation method is randomized, thus capable of producing legitimate outcomes within a finite number of attempts.

\begin{figure}[htp]    
\centering
    \includegraphics[width=.24\textwidth]{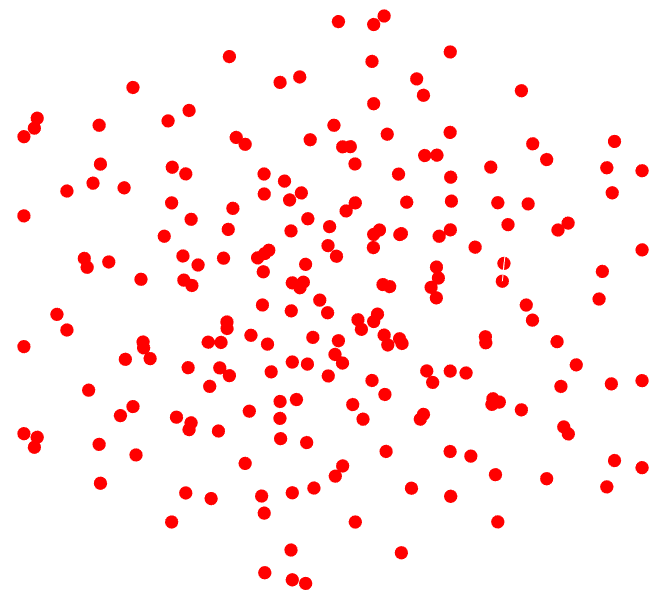}      
    \includegraphics[width=.24\textwidth]{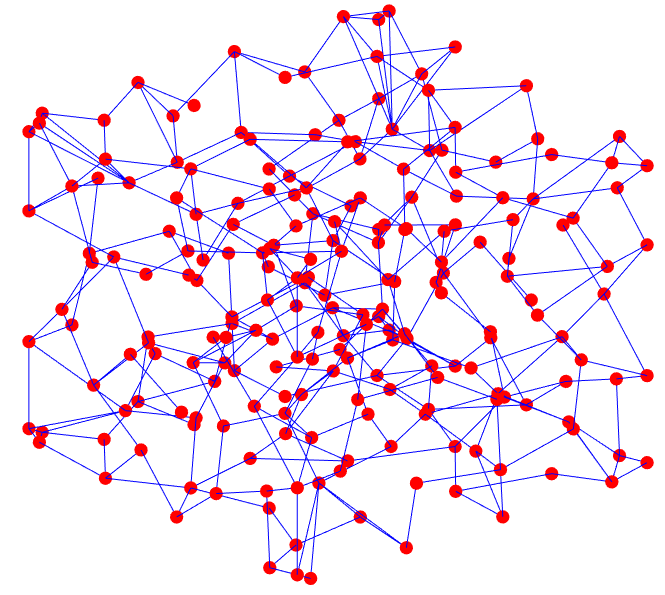}       
    \includegraphics[width=.24\textwidth]{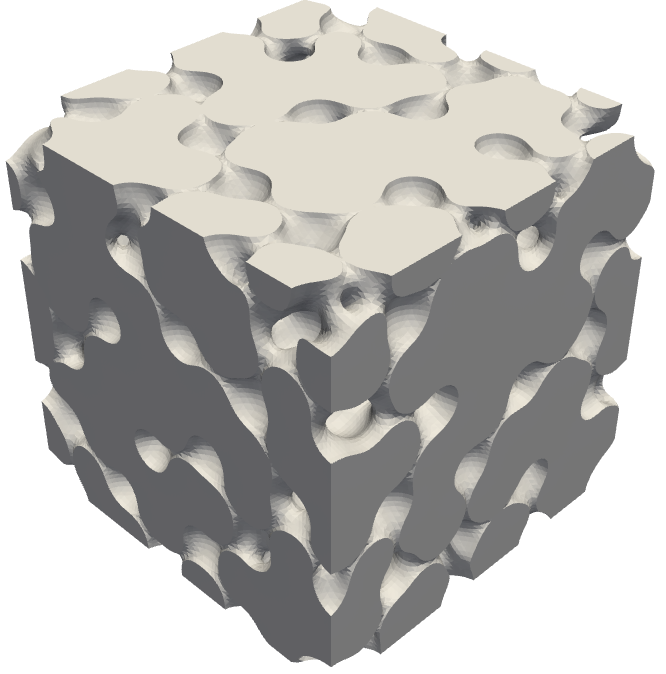}       
    \includegraphics[width=.24\textwidth]{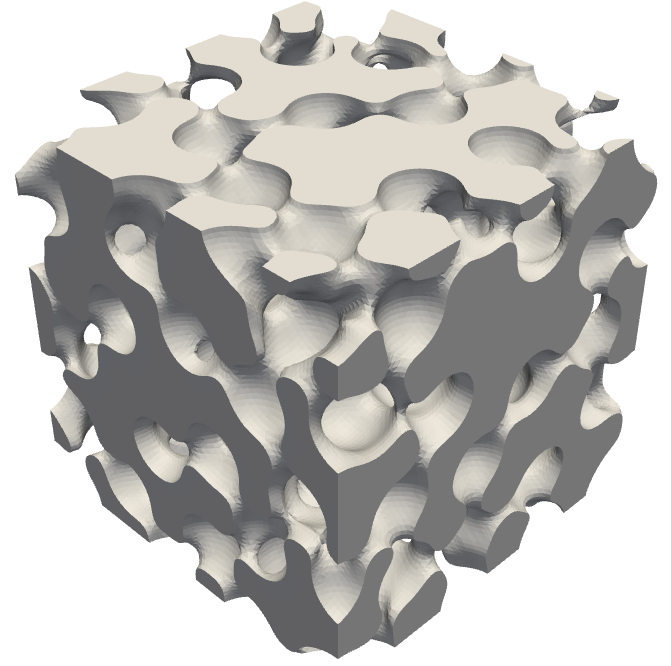}
\centerline{(a)\hspace{0.2\linewidth}(b)\hspace{0.2\linewidth}(c)\hspace{0.2\linewidth}(d)}      
    \caption{The SPPM modeling process. 
    Starting from simultaneous pores sampling on the surface and interior through the blue noise method, represented in the red dots (a), 
tunnel connections (blue edges) between the generated pores are generated based on the modified MST method (b), both the pores and tunnels structures (c) and the SPPM model (d) are reconstructed using the level-set method for implicit expressions.
The number of pores $n=148$,  the number of tunnels is 590.}
    \label{fig:SPPM modeling pipeline}
\end{figure}

Fig.~\ref{fig:SPPM modeling pipeline} delineates the procedure of modeling an SPPM instance.

\subsection{Overall Stochasticity}
\label{subsec:stochastic}
However, tiling only with a same SPPM will induce artificial periodicity at the boundary, which will destroy the stochasticity of the overall porous structure.
Therefore, we design an SPPM tiling rule to optimize the stochasticity of the overall porous structure.
First, SPPM is assembled with both interior and surface parts, so we need to reduce the repetition boundary and change the inner structure to enhance the stochasticity of the structure.

\begin{figure}[htp]
    \centering
    \includegraphics[width=\linewidth]{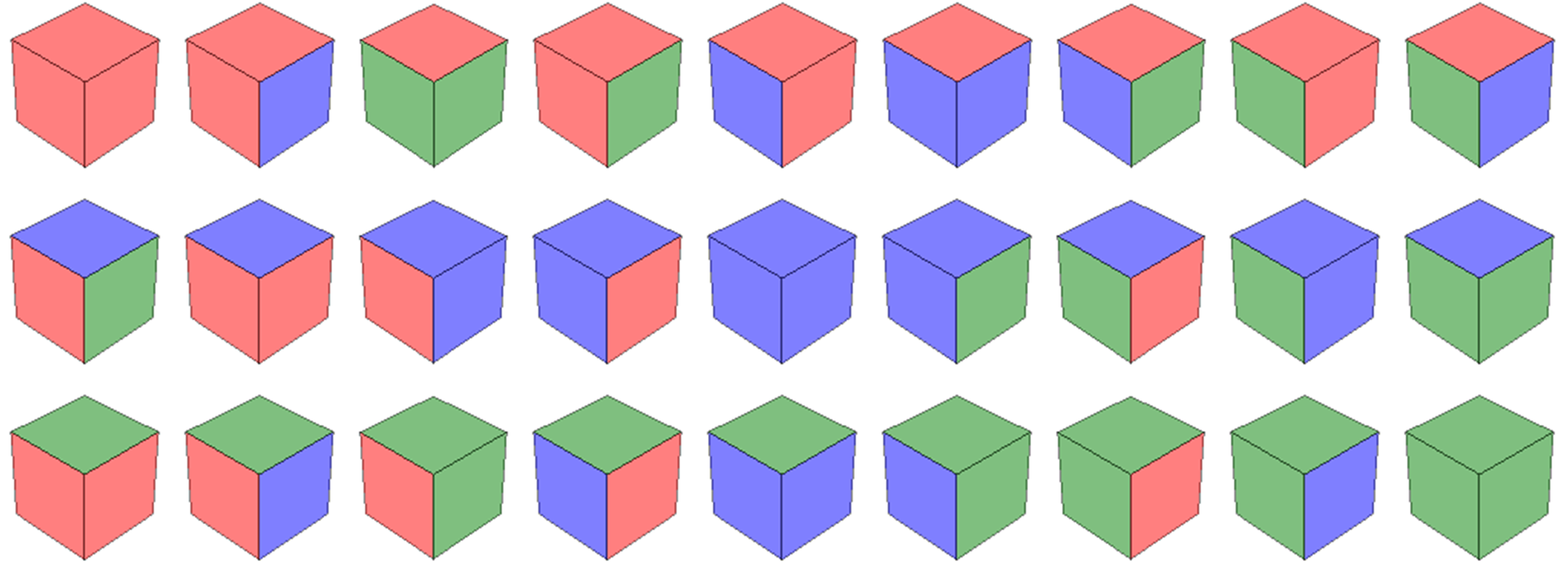}
    \caption{Tile set with 27 tiles defined by 3 colors.}
    \label{fig:tileset}
\end{figure}

We use the idea of Wang tile to design a tile set with different surface and internal pore distributions, as shown in Fig.~\ref{fig:tileset}. Red, blue and green represents three different surfaces with different pore distributions. We iteratively select a tile from this set to place in the space. Specifically, different surfaces are defined as different color codes, combined with different interiors, to generate a microstructure under the condition that the opposite surfaces have the same color code. As shown in Fig.~\ref{fig:tileset}, without considering the rotation, we can generate 27 different microstructure units with three color code surfaces, and they form a tile set. 

We develop two filling rules to determine which microstructure should be placed at each location. 
\emph{Rule 1:} The colors of adjacent boundary surfaces must be consistent. 
\emph{Rule 2:} The placed tiles should be as different as possible from the adjacent ones. 
We quantify the difference by recording a score.

In 3D space, adjacency can be classified into three types: having adjacent point, having adjacent line, and having adjacent surface. We define a function as $\textrm{diff}(A,B)$. If A and B have adjacent surface and have surfaces with different codes then $\textrm{diff}(A,B) = 3$. If A and B have adjacent line and have surfaces with different codes then $\textrm{diff}(A,B) = 2$. If A and B have adjacent point and have surfaces with different codes then $\textrm{diff}(A,B) = 1$.The rest of the cases, $\textrm{diff}(A,B) = 0$. If there is a tile T to be placed, we traverse through its 6 face-adjacent positions, 12 line-adjacent positions and 8 point-adjacent positions to obtain the final score of T which can be represented as

\begin{equation}
 Score =\sum_{i}^{6}\textrm{diff}(F_i,T)+\sum_{i}^{12}\textrm{diff}(L_i,T)+\sum_{i}^{8}\textrm{diff}(P_i,T)
\end{equation}

\begin{figure}[htp]
    \centering
    \includegraphics[width=\linewidth]{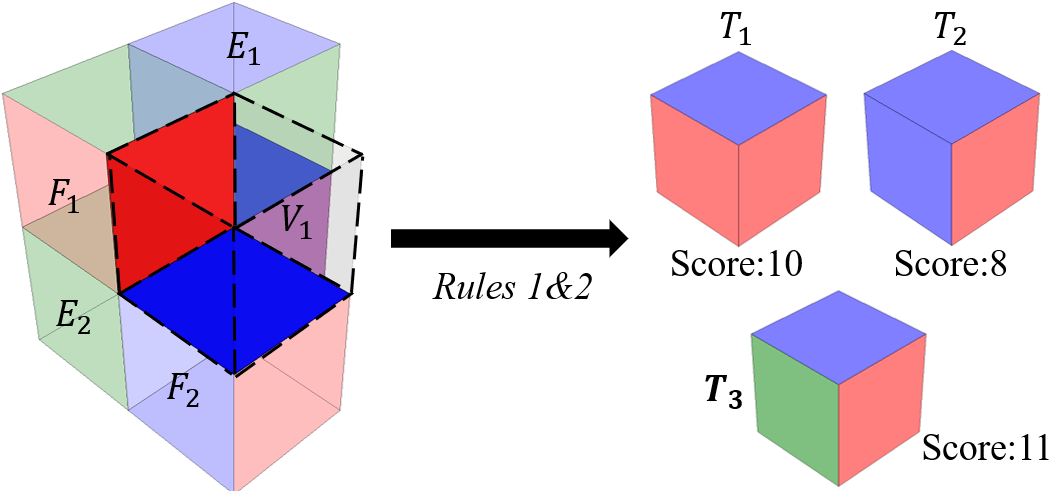}
    \caption{Illustration of the 2 × 2 × 2 structure assembled by the above-proposed tile set. The boundary surfaces constrained by \emph{Rule 1} are marked with an opaque color. The scores calculated by \emph{Rule 2} are shown in the middle. The rightmost is the selected tile.}
    \label{fig:tiling rule}
\end{figure}

Fig.~\ref{fig:tiling rule} demonstrates an example. The already placed tiles are named $F_1$, $F_2$, $L_1$, $L_2$, $P_1$. The tiles that form a face-adjacent relationship with the location to be placed are $F_1$ and $F_2$, the tiles that form a line-adjacent relationship are $L_1$ and $L_2$, and the tiles that form a point-adjacent relationship are $P_1$. The opaque red and blue surfaces represent adjacent surfaces of the tiles to be placed that must have the same color. According to \emph{Rule 1}, we find three tiles that satisfy it ($T_1, T_2, T_3$), as shown in the middle figure. According to \emph{Rule 2}, $T_1$ is the same as $P_1$ and different from the other four tiles, scoring 10. For $T_2$, it is the same as $F_1$ and different from the other four tiles, scoring 8. For $T_3$, it is different from all the other five tiles, scoring 11. Finally, we chose $T_3$ for this location.

\subsection{Porosity Control}
\label{subsec:porosity}

Porosity, defined as the ratio of the non-solid material to the entire material, is an essential feature of porous structures. We adjust the parameters to obtain porous microstructures of the target porosity. 

\begin{figure}[htp]
\centering
    \includegraphics[width=.46\textwidth]{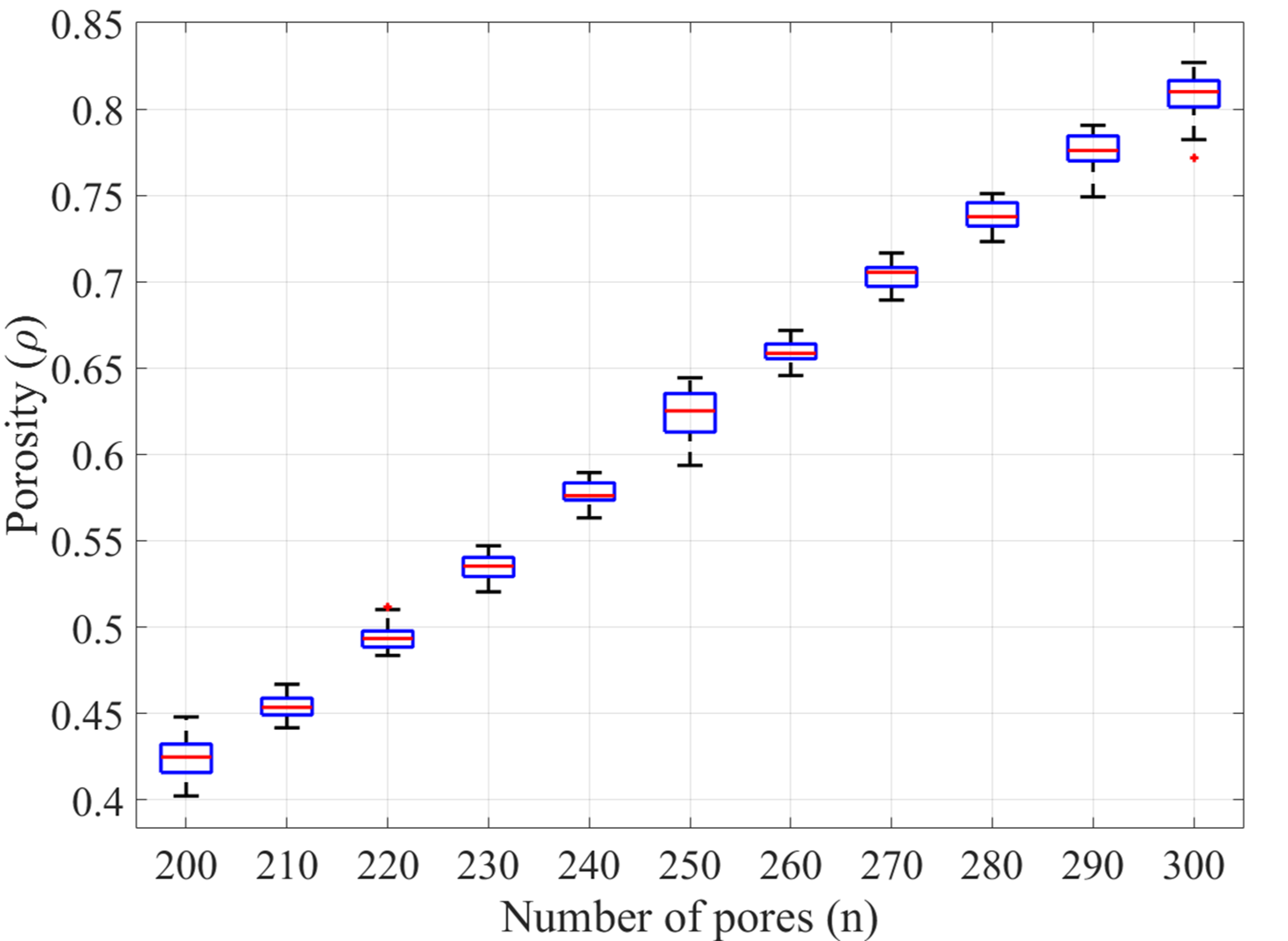}
    \includegraphics[width=.46\textwidth]{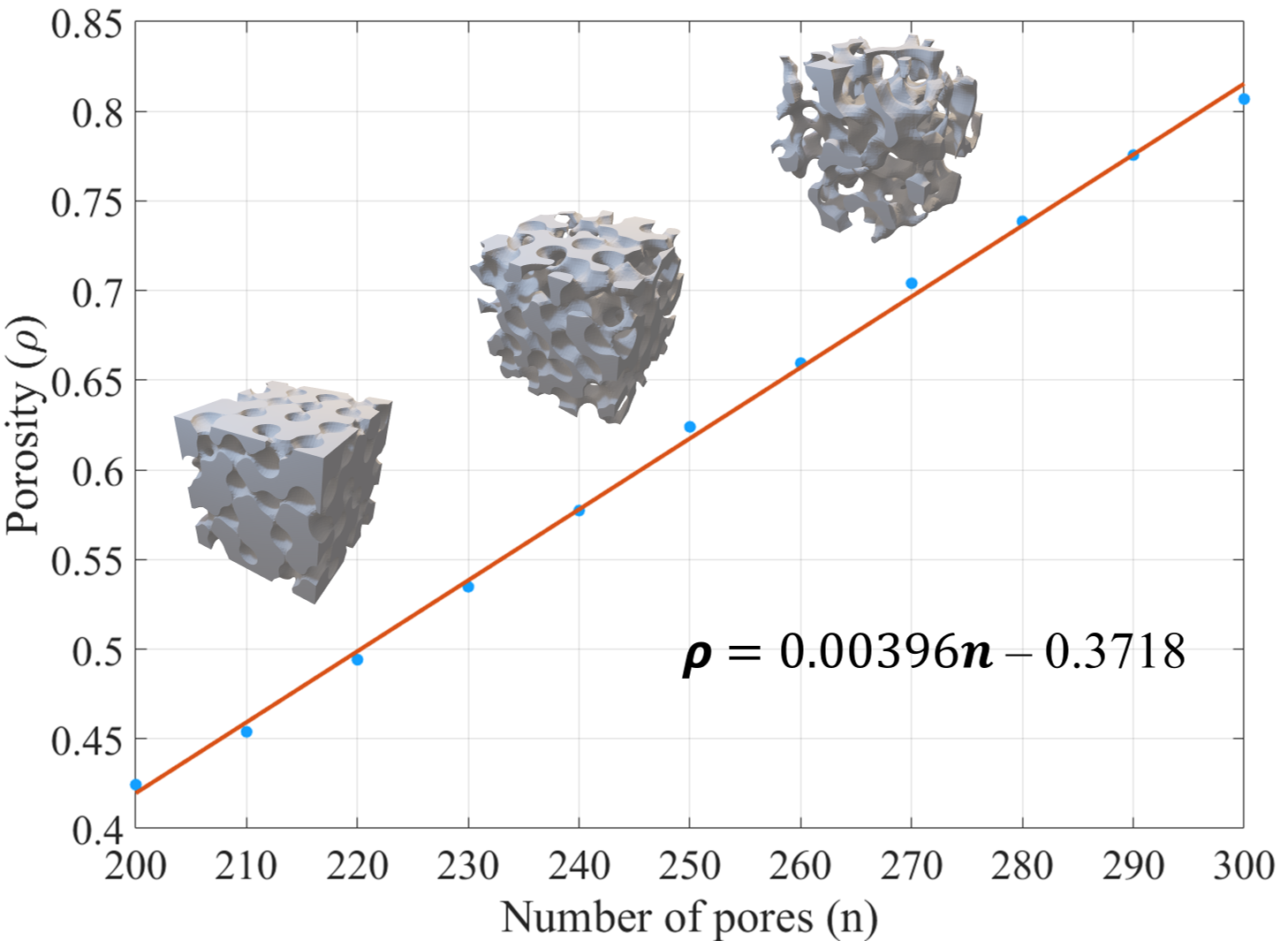}
 \centerline{(a)\hspace{0.4\linewidth}(b)}      
\caption{Analysis of microstructure porosity control. (a) Box plots of number of pores with porosity. 50 samples are generated corresponding to different dart-throwing distances with identical parameters $\omega$ = 30, $\mu$ = 30, $C$ = 0.25, and its porosity are recorded. (b) A fitting model of number of pores ($n$) and average porosity ($\rho$). SPPM with 0.2, 0.5, and 0.8 porosity are also shown in the figure.}
    \label{fig:porosity with n}
\end{figure}

The radius of the pore and the minimum radius of the tunnel are given by the user according to printability considerations, and we can uniquely determine the values of $\omega$ and $\mu$ in the case where the level set value $C$ is determined. 
We then adjust the number of pores $n$ to make the porosity approach the target value within a certain error.
As shown in Fig.~\ref{fig:porosity with n}(a), we set $\omega = 30$, $\mu = 30$, $C = 0.25$, generate 50 samples with the same $n$, and record the porosity of the SPPM.
It can be deduced that the error in porosity of the structures generated with the same $n$ does not exceed $5\%$ in most cases when the pore radius and the minimum tunnel radius are defined. Finally, we fine-tune the interior-to-surface distance $d$ to approximate the given porosity within $0.1\%$ error.
We fitted the curve of $n$ and average $\rho$ with a linear regression model, as shown in Fig.~\ref{fig:porosity with n}(b), for improving the efficiency of searching the target parameters.

\begin{figure}[htp]
\centering
    \begin{minipage}[t]{0.4\linewidth}
        \centering
        \includegraphics[width=\textwidth]{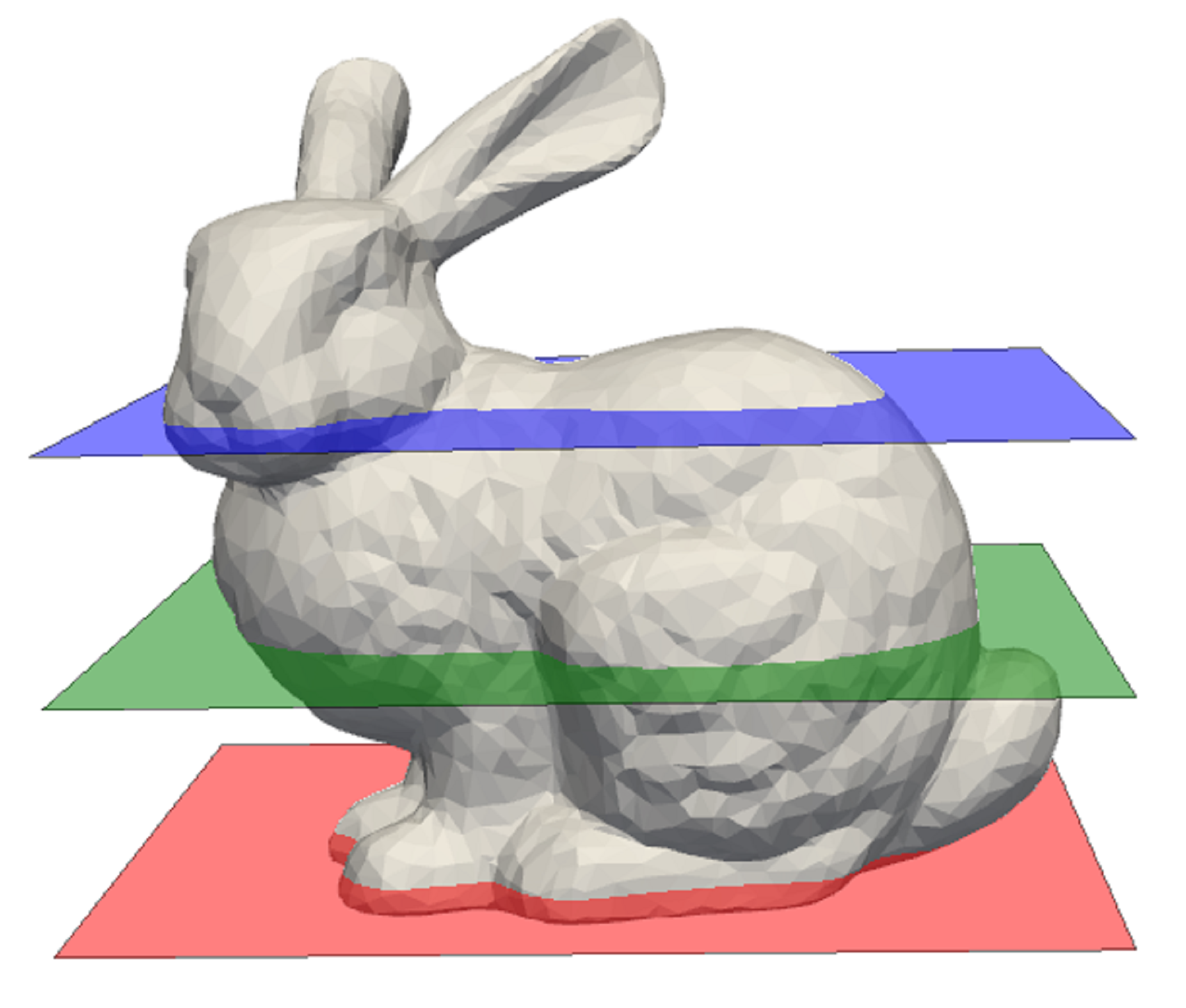}
        \centerline{(a)}
    \end{minipage}%
    \begin{minipage}[t]{0.5\linewidth}
        \centering
        \includegraphics[width=\textwidth]{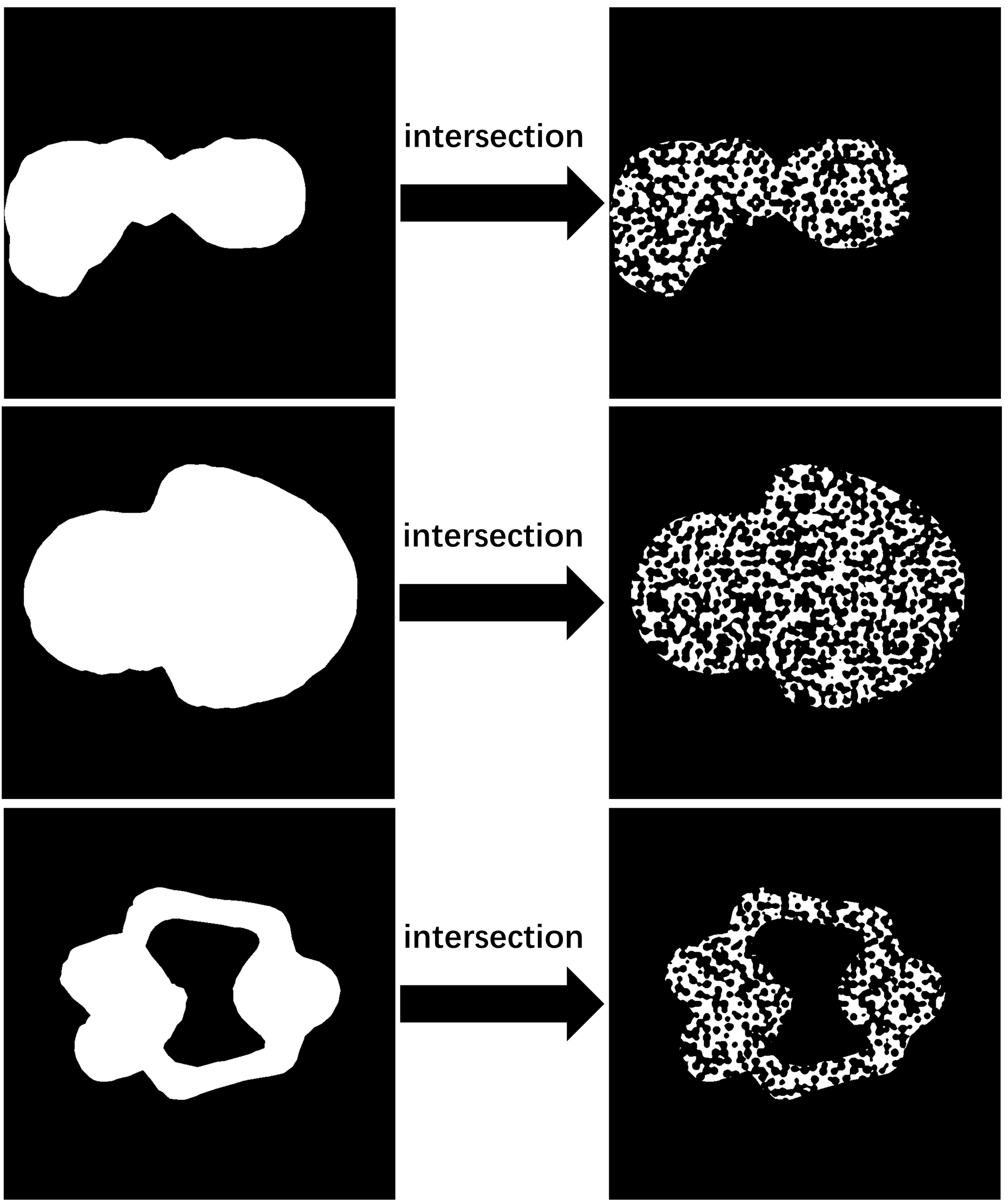}
        \centerline{(b)}
    \end{minipage}
    \caption{(a) The red, green and blue planes represent the three intercepted layers of slices. (b) The results of model slices representing the red, green and blue cross-sections from bottom to top and the microstructure filling results after the intersection operation.}
    \label{fig:Mesh free method}
\end{figure}

\subsection{Mesh-free Slicing}
\label{subsec:meshfree}
To avoid the large amount of time and space consumption associated with reconstructing the mesh, we propose a mesh-free method for generating print results based on implicit fields. 
In brief, we perform Boolean intersection operations on the implicit fields of the model domain and SPPM domain to obtain the slicing results for each layer. 
Take the example of generating a print file in PNG format that is usable by DLP printers. For each location in the modeling space, if the location is within both the model and SPPM, the value is assigned to 255, otherwise the value is assigned to 0. This operation is performed layer by layer for each layer of slices in the modeling domain to obtain the slice file for each layer.
It is worth noting that in this process, the input model shape is an implicit field expression, which can be obtained by calculating the SDF or other methods.
Consequently, the Boolean operation is inside the discrete implicit field rather than the mesh representation to save computation time.
For example, we use the bunny as the model domain, corresponding to the generation of porous microstructures with a 0.6 porosity, using the tiling rule described in Section~\ref{subsec:stochastic} to generate the microstructure domain. 
Finally, the Boolean intersection operation is performed layer by layer.
Fig.~\ref{fig:Mesh free method} shows the results of the three layers of sections intercepted from it and the sections after filling the SPPM.

\section{Results and Discussion}
In this section, we analyze the geometric properties as well as the elastic properties of SPPM. First, we show some SPPMs at the target porosity and the modeling parameters. Also, we show the geometric effect of the stochastic enhanced method. Then we analyzed the elastic properties of SPPMs based on the homogenization method, including Young's modulus, Poisson's ratio, and isotropy. At the same time, we designed physical experiments to evaluate the energy absorption of SPPMs. Finally, we generated two additive manufacturing results that 
show the possibility of generating functional gradient structures and the mesh-free method for efficiently generating print result in arbitrary design domains, respectively.

\begin{figure}[htp]
    \centering
    \includegraphics[width=\linewidth]{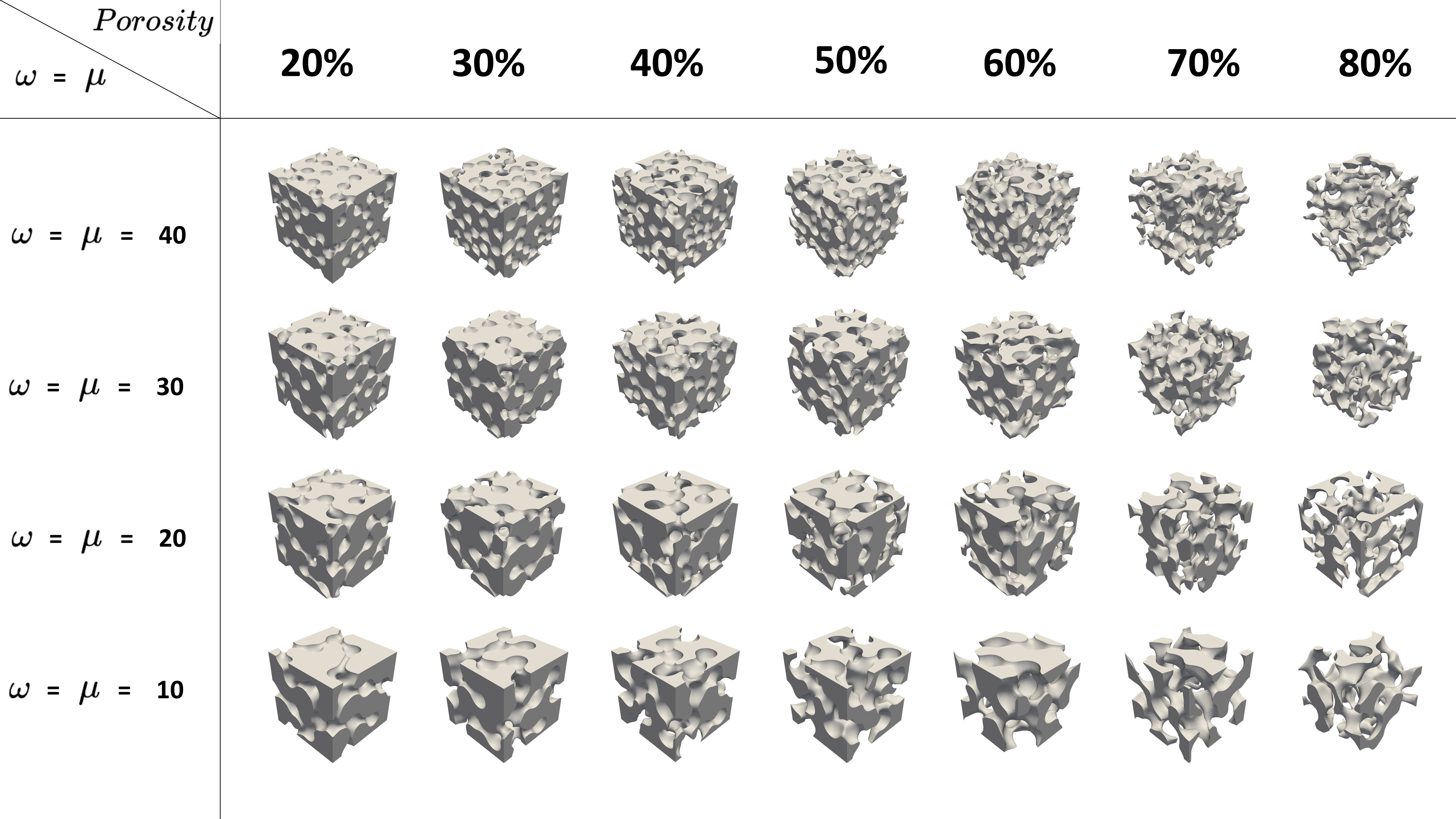}
    \caption{Porous structure of target porosity and target pore and tunnel size. Each structure in the diagram conforms to the periodicity.}
    \label{fig:porous structure of target porosity}
\end{figure}

\subsection{Analysis of geometric properties}

\subsubsection{Porosity}
As described in Section~\ref{subsec:porosity}, the control of porosity can be considered as a search problem with these modeling parameters ($P$, $\theta$). 
After determining the target porosity $\rho$, the number of pores $n$ can be calculated based on the given pore size ($\omega$) and tunnel size ($\mu$). 
The structure of the target porosity is then obtained by fine-tuning the $d$ which is the distance from the surface to the interior. The distance $d$ is only used as a fine-tuning value to approximate the target porosity. If $d$ is decreased, the porosity becomes larger, and if $d$ is increased, the porosity becomes smaller.

Since $\omega$ and $\mu$ are inversely proportional to the size of the pores and tunnels, we ensure that porosity is proportional to the number of pores and proportional to the size of the pores and tunnels.
Fig.~\ref{fig:porous structure of target porosity} lists some SPPMs with target porosity, pore size and tunnel size. It shows a randomly generated microstructure for each specific target. Each result is guaranteed to have periodic boundaries as well as a relatively random distribution of pores.
After our tests, we found that the porosity of our SPPMs is between $20\%$ and $80\%$ to best show the characteristics of the porous structure. 
At porosity less than $20\%$, the elastic characteristics of the microstructure are similar to the base material and the topology is too simple. 
At porosity greater than $80\%$, the solid part shows a tendency to separate and form unconnected islands.

\subsubsection{Microscopic and macroscopic structure}

\begin{figure}[htp]
    \begin{minipage}[t]{\linewidth}
        \centering
        \includegraphics[width=\textwidth]{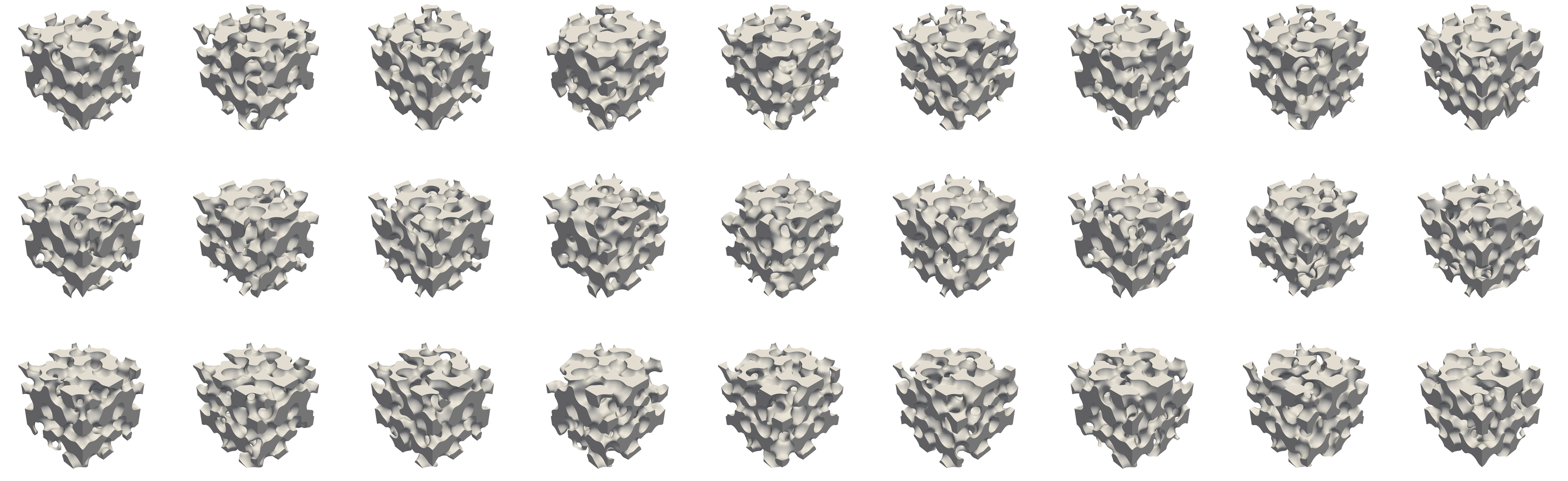}
        \centerline{(a)}
    \end{minipage}%
    
    \begin{minipage}[t]{\linewidth}
        \centering
        \includegraphics[width=\textwidth]{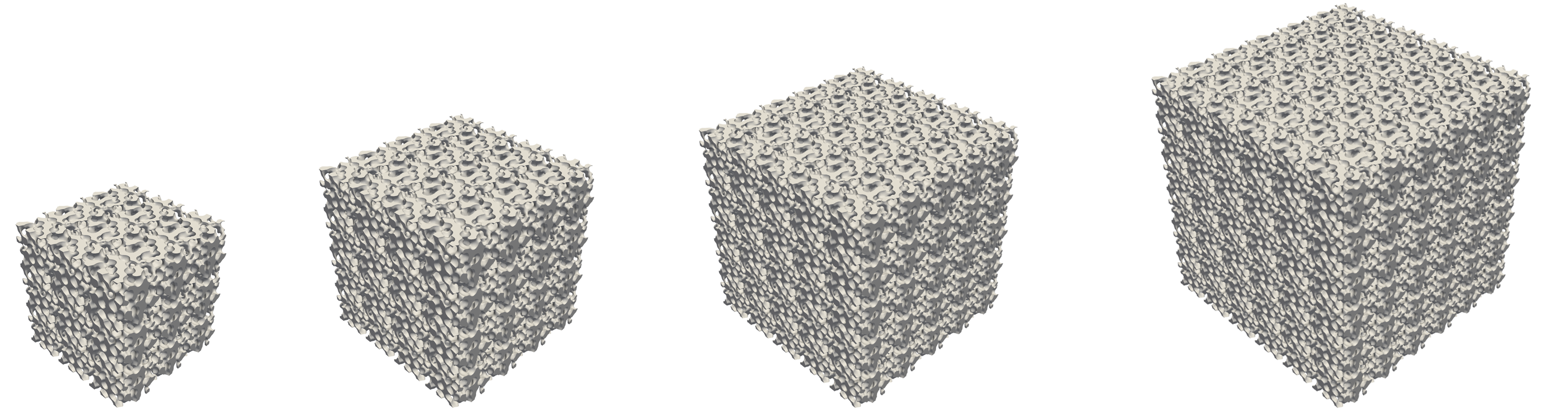}
        \centerline{(b)}
    \end{minipage}
    
     \begin{minipage}[t]{\linewidth}
        \centering
        \includegraphics[width=\textwidth]{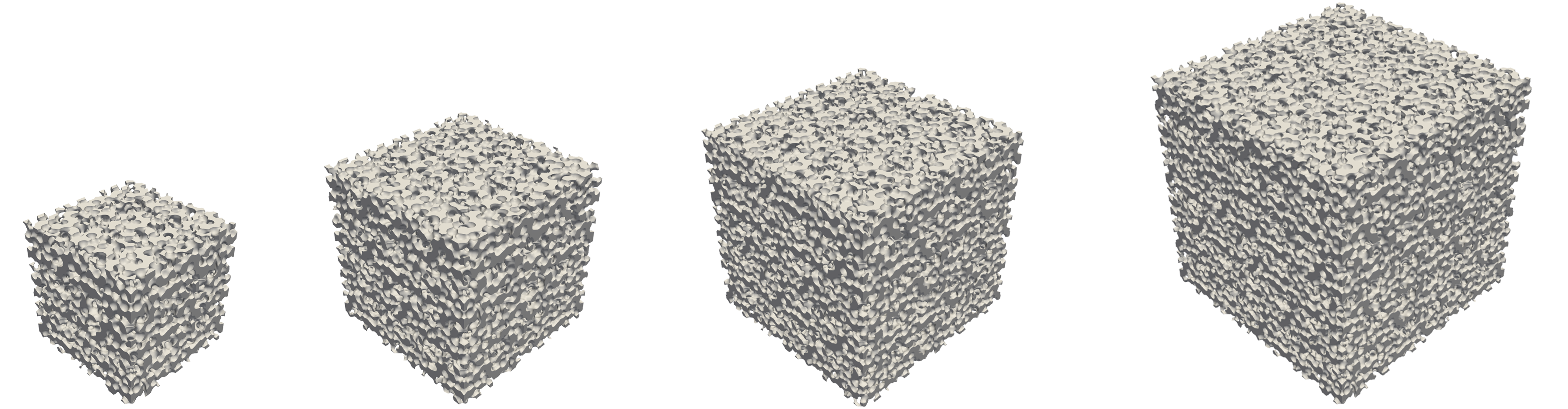}
        \centerline{(c)}
    \end{minipage}
    \caption{Unit cells tilling results. (a) The tiles set was generated using the method described in Section~\ref{subsec:stochastic}. The specific parameters are set as $\omega = \mu = 20$, $\rho = 0.55$. 
    (b) Cell-tiling structure with the same tile from the above tiles set. 
    (c) Cell-tiling Structure with the tiles in (a) and the rules proposed in Section~\ref{subsec:stochastic}. 
    In (b) and (c), the structure is $3\times 3\times 3$, $4\times 4\times 4$, $5\times 5\times 5$, $6\times 6\times 6$ from left to right.}
    \label{fig:periodic and stochastic}
\end{figure}

To verify the effectiveness of our method as well as the visual effect of the stochastic enhancement method.
We generate the tiles set using the method described in Section~\ref{subsec:stochastic} and tile the macroscopic structure using mentioned rules in Section~\ref{subsec:stochastic}.
We set the parameters $\omega = \mu = 20$, $\rho = 0.55$, and the generated tiles setting is shown in Fig.~\ref{fig:periodic and stochastic} (a). 
The porous structure of $3\times 3\times 3$, $4\times 4\times 4$, $5\times 5\times 5$, $6\times 6\times 6$ generated with the stochastic enhancement method is shown in Fig.~\ref{fig:periodic and stochastic} (c).
As a comparison, we used the same SPPM to generate $3\times 3\times 3$, $4\times 4\times 4$, $5\times 5\times 5$, $6\times 6\times 6$ structure as shown in Fig.~\ref{fig:periodic and stochastic} (b). 
Visually we can observe that the structures tilled together with the same microstructures show a clear periodicity. 
In contrast, the structure tilled together with the stochastic enhancement method is more similar to a stochastic structure.

\subsection{Analysis of elastic properties}


\begin{figure}[htp]
    \begin{minipage}[t]{\linewidth}
        \centering
        \includegraphics[width=\textwidth]{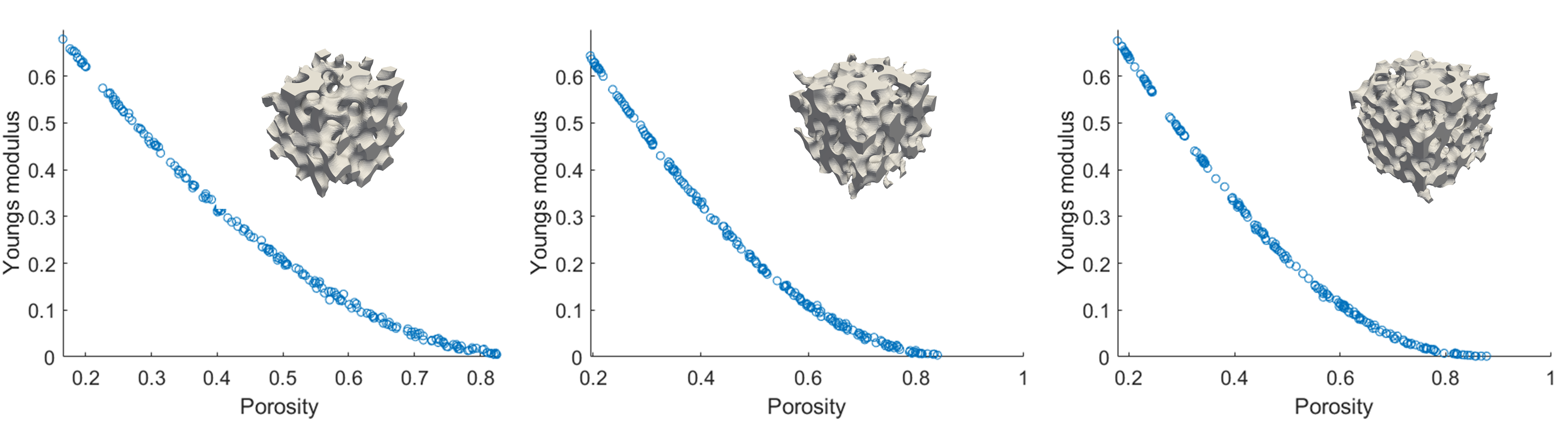}
        \centerline{(a)}
    \end{minipage}%
    
    \begin{minipage}[t]{\linewidth}
        \centering
        \includegraphics[width=\textwidth]{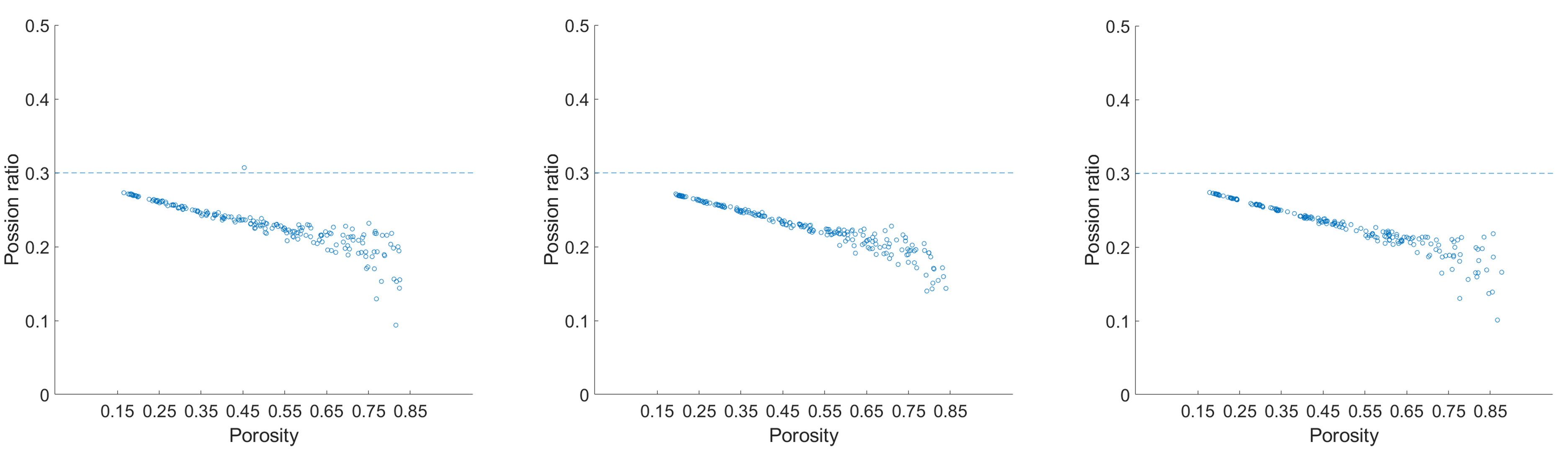}
        \centerline{(b)}
    \end{minipage}
    
     \begin{minipage}[t]{\linewidth}
        \centering
        \includegraphics[width=\textwidth]{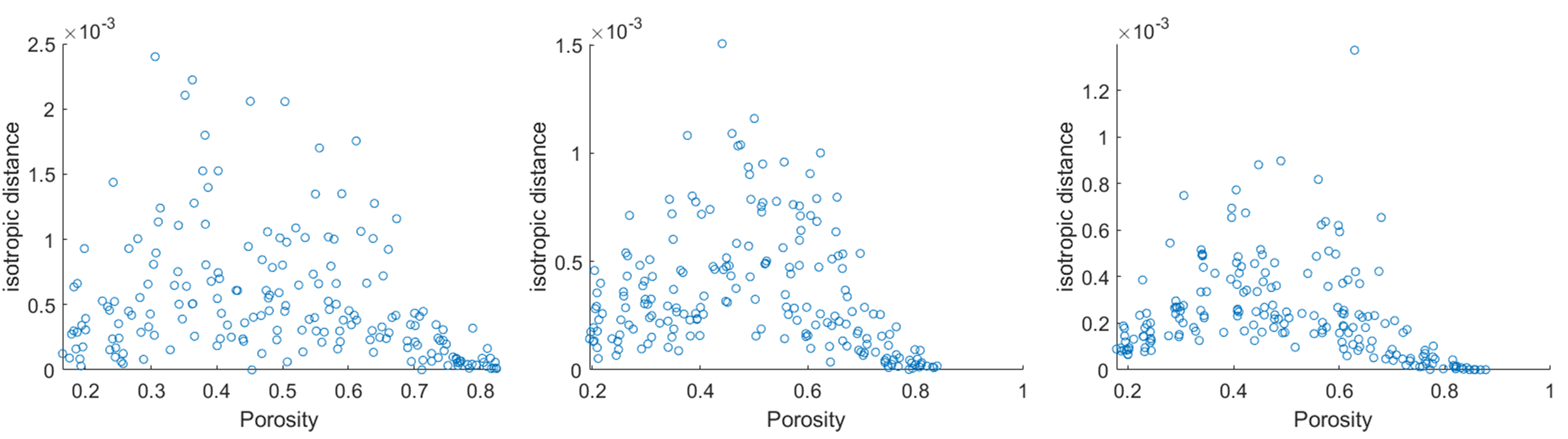}
        \centerline{(c)}
    \end{minipage}
    \caption{Scatter plot of elastic properties analysis of SPPMs. Homogenization is performed at a resolution of $32\times32\times32$. (a) (b) (c) are schematic diagrams of Young's modulus, Poisson's ratio, and isotropy as a function of porosity, pore, and tunnel size in elastic property analysis, respectively. 
    The modeling parameters from left to right are $\omega = \mu = 20$, $\omega = \mu = 30$, $\omega = \mu = 40$, respectively, with pore and tunnel size from large to small.}
    \label{fig:cell_elastic_properties}
\end{figure}

Since the SPPMs are periodic, we can perform homogenization analysis on a base volume cell to obtain the elastic characteristics of the macroscopic structure. 
By homogenization analysis of the microstructure with different parameters $(\omega, \mu, \rho)$, we reconstruct the potential relationship between these parameters and the elastic characteristics (Young's modulus, Poisson's ratio, and isotropic deviation). 
The graph resulting from homogenization analysis is shown in Fig.~\ref{fig:cell_elastic_properties}. The base material properties are set to Young's modulus of 1.0 and Poisson's ratio of 0.3. 

Experimentally we obtain only very small deviations in Young's modulus along the different directions for SPPMs, so we uniformly use Young's modulus along the X-axis, and Poisson's ratio that corresponds to a contraction in direction Y when an extension is applied in direction X.

For the isotropic evaluation of the structure, we calculate the deviation of the tensor of the SPPM from a fully isotropic tensor \cite{Martinez2016}. 
The tensor of a fully isotropic material is denoted as $C^{I}(E,\nu)$, where $E$ is Young's modulus and $\nu$ is Poisson's ratio.

\begin{equation}
\hat{E}  = \frac{E}{(1-2\nu)(1+\nu)}, ~~ G = \frac{E}{2(1+\nu)}
\end{equation}

\begin{equation}
C^{I}(E,\nu) = \begin{pmatrix}
 \hat{E}(1-\nu)  & \hat{E}\nu & \hat{E}\nu &  0&  0&0 \\
  \hat{E}\nu&  \hat{E}(1-\nu) & \hat{E}\nu &0  & 0 &0 \\
 \hat{E}\nu & \hat{E}\nu &  \hat{E}(1-\nu) & 0 & 0 &0 \\
  0& 0 &0  & G & 0 & 0\\
 0 &0  &0  & 0 & G & 0\\
  0& 0 & 0 & 0 &  0&G
\end{pmatrix} 
\end{equation}

We homogenize the generated microstructure and calculate the deviation $\xi (C^{H} )$ of the calculated tensor from the fully isotropic tensor as the distance of the structure from isotropy.
\begin{equation}
\xi (C^{H} ) = \min_{E,\nu} \left \| C^{I}(E,\nu)-C^{H} \right \|^{2} _{F}
\end{equation}
Where $ 0 < E< E_{M}$, $-1<\nu<0.5$, and $E_{M}$ is the upper bound of Young’s modulus of the solid base material. 
We use a nonlinear optimizer to solve this minimization problem.

\begin{figure}[htp]
    \begin{minipage}[t]{\linewidth}
        \centering
        \includegraphics[width=\textwidth]{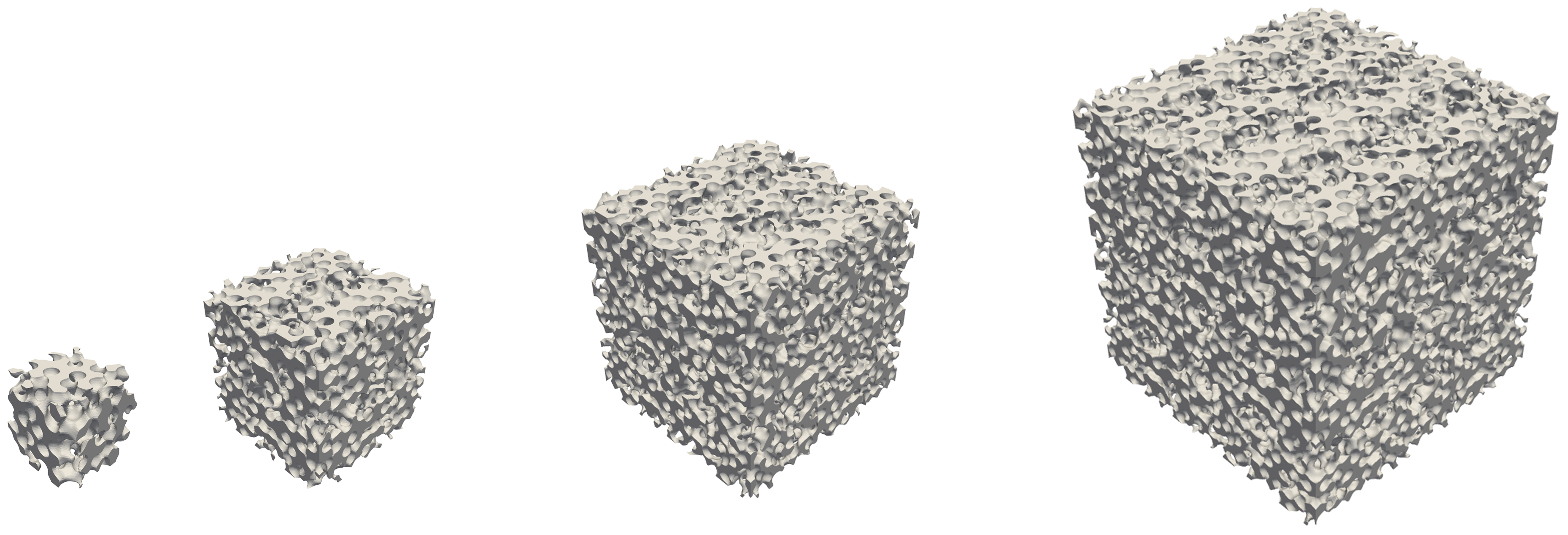}
        \centerline{(a)}
    \end{minipage}%

    \begin{minipage}[t]{0.333\linewidth}
        \centering
        \includegraphics[width=\textwidth]{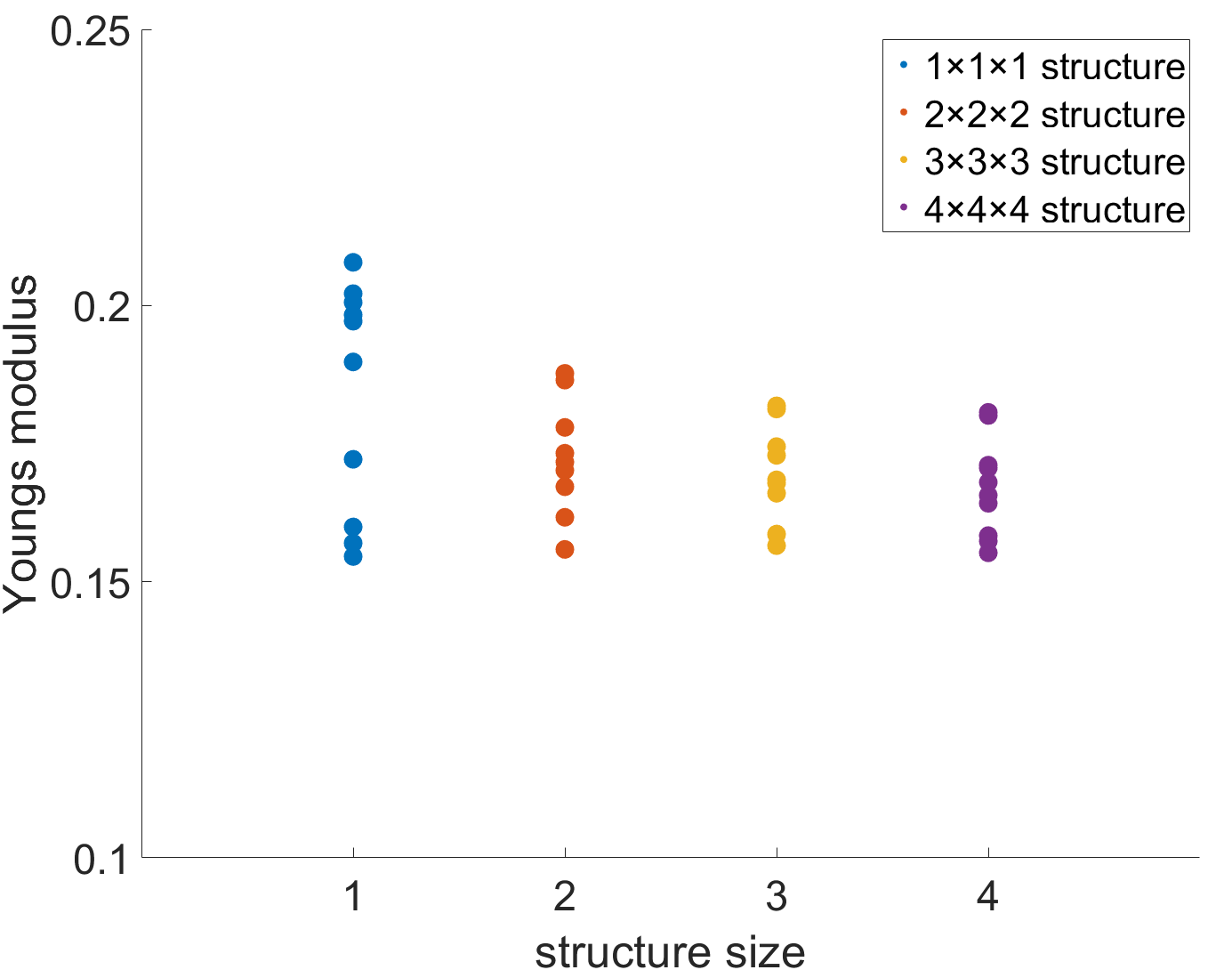}
        \centerline{(b)}
    \end{minipage}%
    \begin{minipage}[t]{0.333\linewidth}
        \centering
        \includegraphics[width=\textwidth]{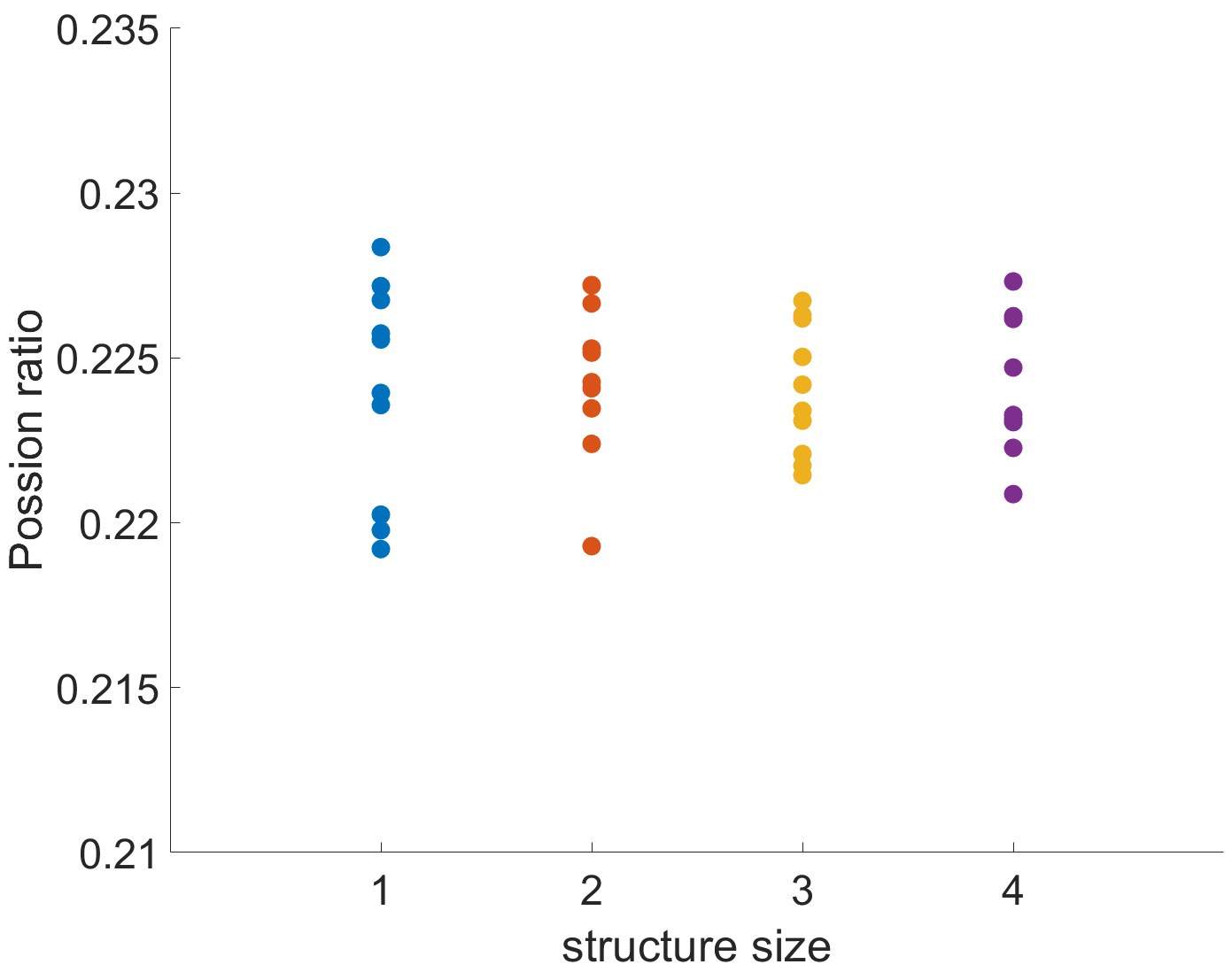}
        \centerline{(c)}
    \end{minipage}%
    \begin{minipage}[t]{0.333\linewidth}
        \centering
        \includegraphics[width=\textwidth]{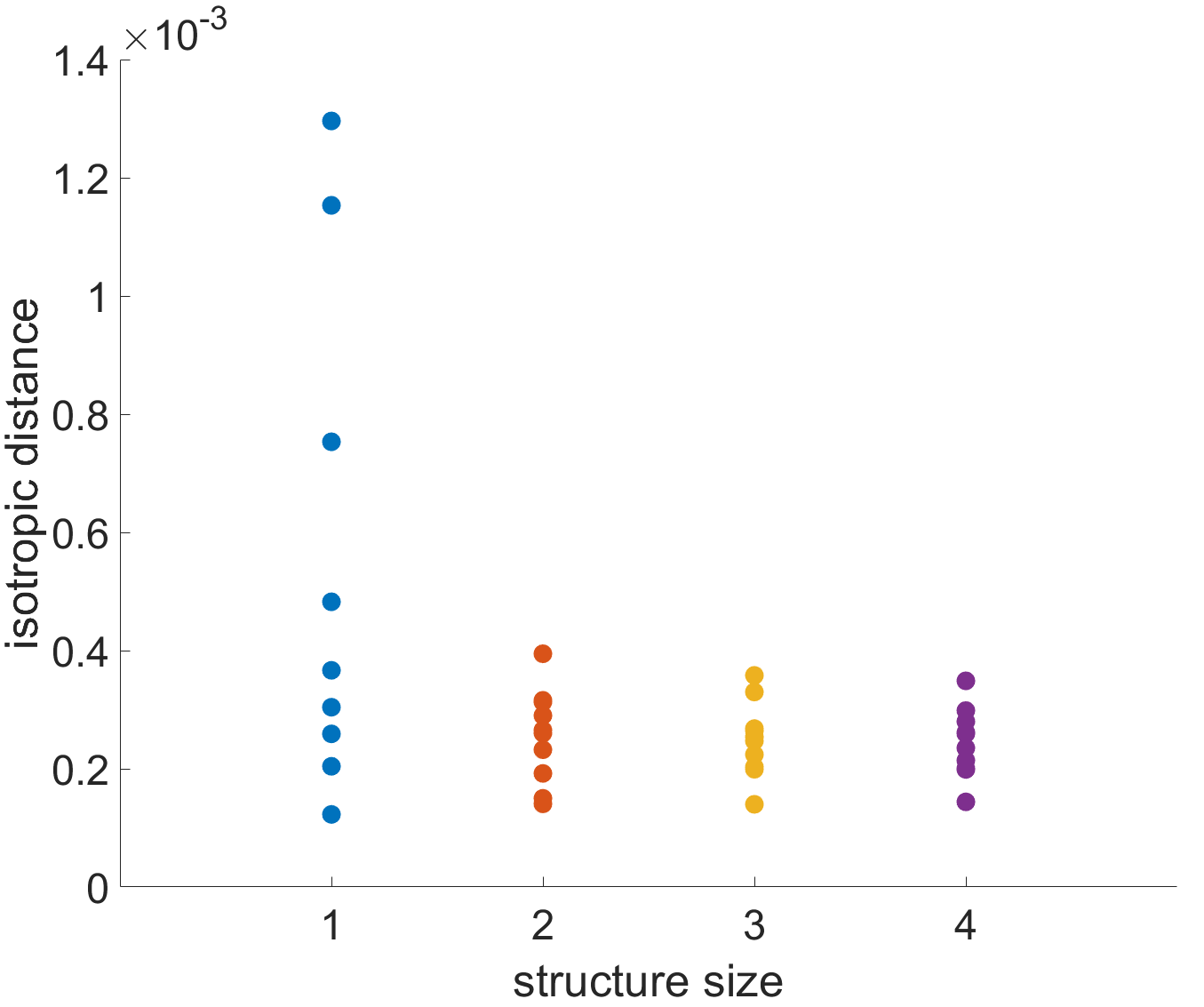}
        \centerline{(d)}
    \end{minipage}
    \caption{(a) Geometric mesh diagrams of porous structures at different sizes of design domains. The number of microstructures contained from left to right is $1\times1\times1$, $2\times2\times2$, $3\times3\times3$, $4\times4\times4$. The resolution of homogenization is 32, 64, 96, 128 respectively. 10 samples are generated for each design domain, (b) (c) (d) corresponding to Young's modulus, Poisson's ratio, and isotropic deviation in elastic properties, respectively.}
    \label{fig:elastic_properties_different_sizes_of_structures}
\end{figure}

Fig.~\ref{fig:cell_elastic_properties} (a) shows the variation of Young's modulus with porosity for different pore and tunnel sizes. 
We can infer that Young's modulus of SPPMs is directly related to the porosity, and there is a continuum of coverage of Young's modulus at different porosities. 
Also, when the porosity is fixed, changing the size of the pores and tunnels has less effect on Young's modulus. 
Therefore we can uniquely determine the structure at a target Young's modulus based on the modeling parameter $(\omega, \mu, \rho)$.
Fig.~\ref{fig:cell_elastic_properties} (b) shows the variation of Poisson's ratio with porosity for different pore and tunnel sizes. 
As can be seen that the Poisson's ratio shows a small decrease with the decrease of Young's modulus and remains similar to the Poisson's ratio of the base material.
Fig.~\ref{fig:cell_elastic_properties} (c) shows the variation of isotropic deviation $\xi (C^{H})$ with porosity for different pore and tunnel sizes. It can be seen that our structures exhibit well isotropy under different modeling parameters.

We then analyzed the elastic properties of the macroscopic structures modeled using the stochastic enhancement method proposed in Section~\ref{subsec:stochastic}. 
As shown in Fig.~\ref{fig:elastic_properties_different_sizes_of_structures} (a), the modeling space increases sequentially from left to right, and the numbers of microstructures included are $1\times1\times1$, $2\times2\times2$, $3\times3\times3$, $4\times4\times4$. 
We randomly generated 10 samples for each modeling space, Fig.~\ref{fig:elastic_properties_different_sizes_of_structures} (b) (c) (d) shows their Young's modulus, Poisson's ratio, and isotropic deviation. 
As can be seen that the elastic properties of the structure show convergence as the modeling space increases. 
Besides, as the modeling space increases, the stochastic enhancement method has an enhanced effect on the isotropy of the overall structure.

\subsection{Energy Absorption Experiments}

\begin{figure}[htp]
    \centering
    \includegraphics[width=\linewidth]{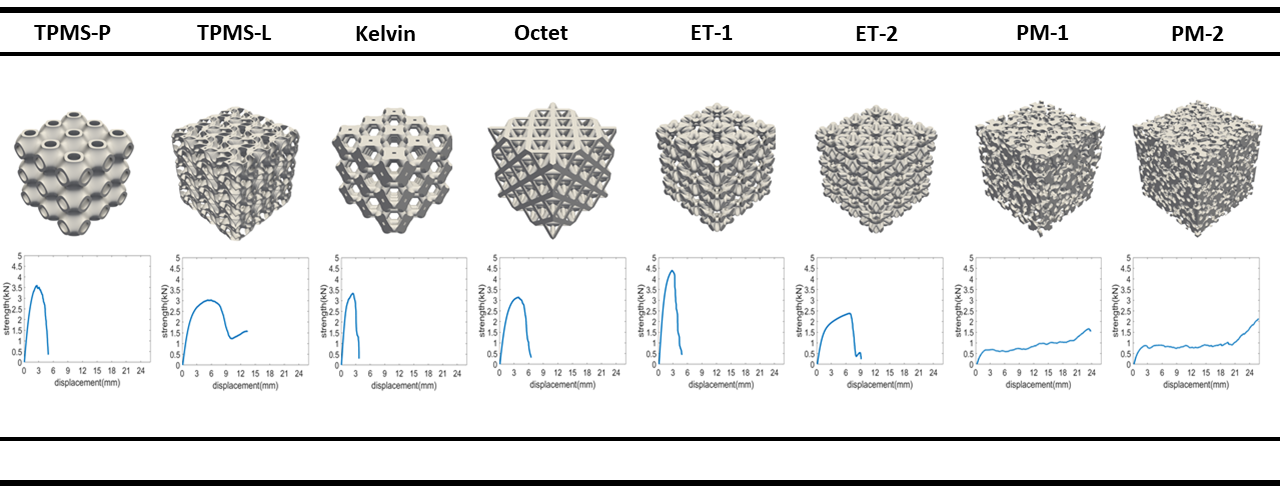}
    \caption{The middle row is $3\times3\times3$ structure and corresponding displacement-force curve. The bottom row shows the total absorption energy before plastic deformation. The volume fraction of the microstructure in the figure is 0.3. The modeling parameters for PM-1 are $\omega = \mu = 20$, for PM-2 are $\omega = \mu = 30$.}
    \label{fig:Energy absorbtion}
\end{figure}

\begin{table}[ht]
\caption{Comparison of effective energy absorption ability of microstructures} 
\label{tab:Curve analysis}
\resizebox{\linewidth}{!}{
\begin{tabular}{cccc}
\toprule 
  Type& Compressive yield strength (MP)
 & Effective displacement (mm)
 & Absorption (J) \\
\midrule 
 TPMS-P &1.7765 &2.6396 &6.065\\
 TPMS-L &1.4936 &5.2674 &16.11\\
 Kelvin &1.6477 &2.3844 &8.357\\
 Octet &1.5488  &3.9342 &12.39\\
 ET-1 &2.1695 &2.8259 &8.753\\
 ET-2 &1.1789 &6.7259 &6.236\\
 SPPM-1 &0.5452 &20.2926 &20.93\\
 SPPM-2 &0.4433 &20.3177 &16.73\\
\bottomrule 
\end{tabular}
}
\end{table}

To investigate the energy absorption characteristic of SPPM, we conducted a uniaxial compression test according to the standard GB/T8813-2008. The base material is DM-12 which Young's modulus is 1404 MPa, Poisson's ratio is 0.45, and tensile strength is 29 MPa. The manufacturing size of each structure is 45mm*45mm*45mm and contains $3\times3\times3$ microstructure units. The loading speed of the testing machine is set to 5mm/min. In the test, the structure undergoes roughly three stages, namely the elastic deformation stage, the yielding stage and the plastic deformation stage. We use the total energy absorption in the elastic deformation stage and yielding stage to measure the energy absorption capacity of this structure and the displacement of the structure in these two stages is recorded as the effective displacement. We calculate the total effective energy absorption of the structure by using the equation. \begin{equation}
    W = Fs\cos \alpha, 
\end{equation}

We conducted comparative experiments using six common microstructures as shown in Fig.~\ref{fig:Energy absorbtion}. They are TPMS-P, TPMS-L, Kelvin, Octet, and two types of Elastic Textures (ET). We performed uniaxial compression tests on two SPPMs with different pore sizes. Their parameters are $\omega = \mu = 20$, $\omega = \mu = 30$ respectively.  Fig. \ref{fig:Energy absorbtion} shows the displacement-force curves of the eight microstructures. As can be seen, for the six compared structures, their yield stage is very short and the yield strength is the force at the highest point of the curve divided by the area under force. The effective displacement is the displacement reached at the yield strength. In the case of SPPMs, they have relatively long yield phase but a relatively small yield strength. Table~\ref{tab:Curve analysis} shows the yield strength, effective displacement, and total effective energy absorption for the eight microstructures.

\subsection{Additive Manufacturing}
We test our algorithm on a Windows 10 PC with a 2.5GHZ Core CPU and 16GB RAM, and use a DLP (Digital Light Processing) printer for manufacturing. 

\begin{figure}[htp]
    \begin{minipage}[t]{0.5\linewidth}
        \centering
        \includegraphics[width=0.9\textwidth]{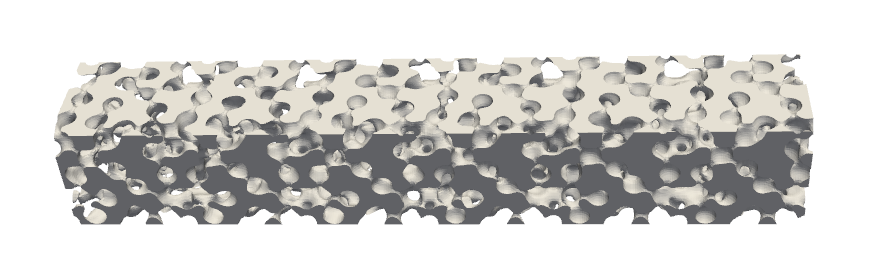}
        \centerline{(a)}
    \end{minipage}%
    \begin{minipage}[t]{0.5\linewidth}
        \centering
        \includegraphics[width=0.9\textwidth]{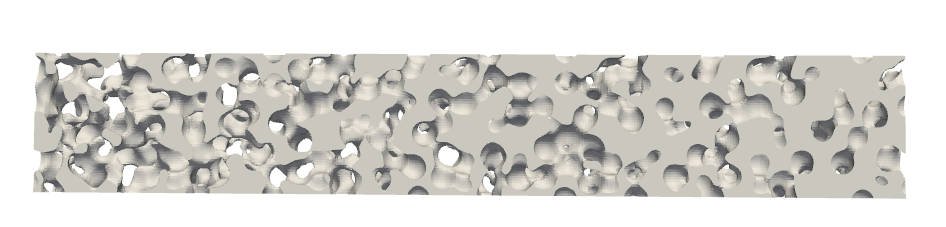}
        \centerline{(b)}
    \end{minipage}%
    
    \begin{minipage}[t]{0.5\linewidth}
        \centering
        \includegraphics[width=0.9\textwidth]{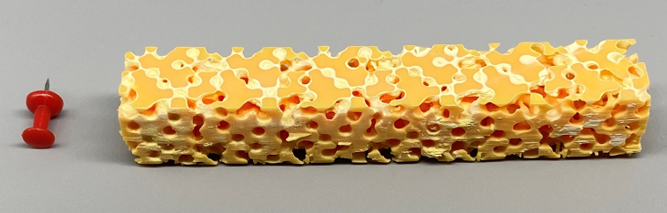}
        \centerline{(c)}
    \end{minipage}%
    \begin{minipage}[t]{0.5\linewidth}
        \centering
        \includegraphics[width=0.9\textwidth]{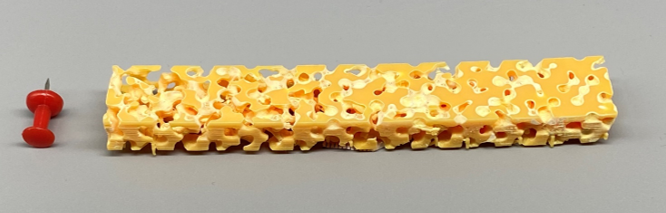}
        \centerline{(d)}
    \end{minipage}
    \caption{Functional gradient structure of overall structure mesh (a), section mesh (b), overall structure manufacturing result diagram (c), section manufacturing result diagram (d). The length, width and height of the model in (c) are 120 mm, 20 mm and 20 mm respectively, and in (d) are 120 mm, 20 mm and 10 mm respectively. The dimension of one single microstructure is approximately 20 mm.}
    \label{fig:Grading_bar}
\end{figure}

First, to verify that our method could model the structure of functional gradients, we generated a bar containing six SPPMs with pore density ranging from large to small. 
Fig.~\ref{fig:Grading_bar} (a) and Fig.~\ref{fig:Grading_bar} (b) are the overall structure mesh and section mesh of the functional gradient structure, respectively. Fig.~\ref{fig:Grading_bar} (c) and Fig.~\ref{fig:Grading_bar} (d) are the overall structure manufacturing results and section manufacturing results of the functional gradient structure, respectively. 
The limitation of fabricating functional gradient structures with our method is that the interior of the structure can vary with the gradient, but the boundaries appear repetitive.

\begin{table}[ht]
  \caption{Comparison of the efficiency of mesh and mesh-free modeling methods}
  \label{tab:Comparison of efficiency}
\resizebox{\linewidth}{!}{
\begin{tabular}{*{5}{c}}

  \toprule
  \multirow{2}*{Number of units} & \multicolumn{2}{c}{Storage consumption (MB)} & \multicolumn{2}{c}{Time consumption (ms)} \\
  \cmidrule(lr){2-3}\cmidrule(lr){4-5}
  & Mesh & Mesh-free & Mesh & Mesh-free  \\
  \midrule
  27 & 95.4 & 1.48 & 3082 & 811 \\
  216 & 801 & 5.49 & 54276 & 1955 \\
  512 & 1945 & 10.5 & 343623 & 3274 \\
  \bottomrule
\end{tabular}
}
\end{table}

To test the efficiency of the mesh-free modeling method in time and storage, we design an experiment to compare the efficiency of the mesh and mesh-free modeling methods. We generate model files in OBJ format as a result of the mesh modeling method, with storage space of the size of the OBJ model. The mesh generation with MC (Machine Cube) method is performed from the initial parameter setting to the generation of OBJ file as the time consumption of the mesh modeling method. The resolution of each microstructure is set to $32\times32\times32$. For the mesh-free method, we use the generated print file as the result of the method, i.e., a set of PNG (Portable Network Graphics) images for the DLP printer. The storage consumption is the total space occupied by the PNG images collection, and the time consumption is the total time from the initial parameter setting to the generation of the PNG image collection. 
We test three cube structures containing 27, 216, and 512 units, respectively. The comparison of the efficiency of mesh and mesh-free modeling methods is shown in Table \ref{tab:Comparison of efficiency}. It can be seen that the mesh-free modeling method is much less time and space-consuming than the mesh modeling method.

\begin{figure}[htp]
    \begin{minipage}[t]{0.333\linewidth}
        \centering
        \includegraphics[width=0.9\textwidth]{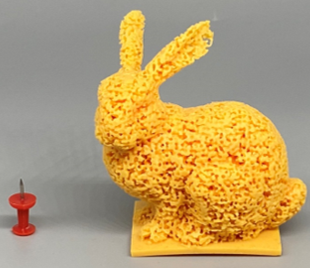}
        \centerline{(a)}
    \end{minipage}%
    \begin{minipage}[t]{0.333\linewidth}
        \centering
        \includegraphics[width=0.9\textwidth]{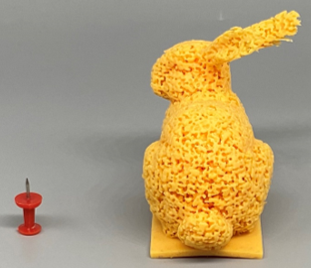}
        \centerline{(b)}
    \end{minipage}%
    \begin{minipage}[t]{0.333\linewidth}
        \centering
        \includegraphics[width=0.9\textwidth]{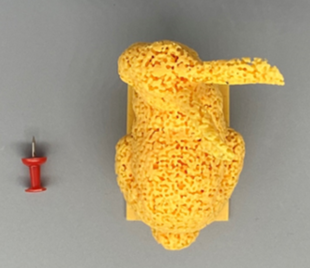}
        \centerline{(c)}
    \end{minipage}%
    \caption{A bunny model tiled with SPPMs with $70\%$ porosity. (a) (b) (c) is the front view, side view and top view, respectively. The length, width and height of the model are 59.85 mm, 46.43 mm, 60 mm. The dimension of one single microstructure is approximately 7.5 mm.}
    \label{fig:Manufacture_bunny}
\end{figure}

Then, to test our method on arbitrary shapes are satisfied with the printing requirements. We tile a bunny model with mesh-free modeling methods. The fabrication result shown in Fig.~\ref{fig:Manufacture_bunny} contains approximately 150 SPPMs at the same time the dimension of every microstructure is approximately 7.5 mm. Every SPPMs have a porosity of 0.7 and a resolution of $100\times100\times100$. The mesh reconstruction of this structure is very time consuming and memory intensive, which is difficult to implement on our PC.

\section{Conclusions}
\label{sec:conclusion}
This study has elucidated our method for modeling and performance analysis of Stochastic Porous Periodic Materials (SPPM). Our approach employs a probability-based stochastic enhancement algorithm that manipulates the basic volume unit to generate SPPM models with varying pore sizes and shapes. 
It has greatly enhanced the efficiency of generating extensive stochastic microstructures. By integrating the benefits of stochastic and periodic microstructures, we have successfully achieved a high level of isotropy in physical properties.
Further assessment of these models revealed excellent performance in terms of elastic modulus, Poisson's ratio, and isotropic properties.

However, our research is not without limitations. First, the experimental part of this study is confined to a single material, which may not fully reflect the behavior of different materials in SPPM modeling. Secondly, we currently employ the same boundary constraints, which could potentially impact the performance of the model in specific application scenarios. Furthermore, the current structure at the boundary is C0 continuous, which might cause local smoothness, an issue we aim to address at the algorithmic level in the future.

Further optimization of SPPM design to accommodate a wider range of application scenarios and material demands. We plan to explore more boundary conditions and other parameter settings to enhance the versatility of SPPM.
Adaptively connect microstructural units with different boundaries when generating the graded structure. This approach could potentially improve the overall performance of the generated structures.
Broadening the experimental scope by testing the performance of SPPM on different types of materials. This would enable a more comprehensive understanding of SPPM's performance in various environments.
Exploring the application of SPPM in other fields such as acoustics, thermodynamics, and electromagnetics. We believe that SPPM, owing to its unique structural properties, has the potential to play a significant role in these fields.
To further enhance model accuracy, we aim to incorporate more practical factors into the model, such as material fatigue performance, temperature effects, etc.
Designing a functionally graded structure using our algorithm and investigating its performance across various applications.

In conclusion, this study provides an effective method for SPPM modeling and performance analysis. Despite certain limitations, we are confident that with continuous optimization and improvement, our method holds promise for significant contributions to material science, manufacturing engineering, and other related fields.
\bibliographystyle{elsarticle-num} 
\bibliography{main}
\end{document}